\documentclass[aps,
prd,
english,preprintnumbers,nofootinbib,
twocolumn]{revtex4-1}

\usepackage{amsfonts,amsmath,amssymb}
\usepackage{graphicx}
\usepackage[utf8]{inputenc}
\usepackage{hyperref}
\usepackage{babel}
\usepackage{enumitem}

\newcommand\e{{\rm e}}

\newcommand\be{\begin{equation}}
\newcommand\ee{\end{equation}}
\newcommand\bea{\begin{eqnarray}}
\newcommand\eea{\end{eqnarray}}

\begin{document}

\def\rhoo{\rho_{_0}\!} 
\def\rhooo{\rho_{_{0,0}}\!} 

\begin{flushright}
\phantom{
{\tt arXiv:2006.$\_\_\_\_$}
}
\end{flushright}

{\flushleft\vskip-1.4cm\vbox{\includegraphics[width=1.15in]{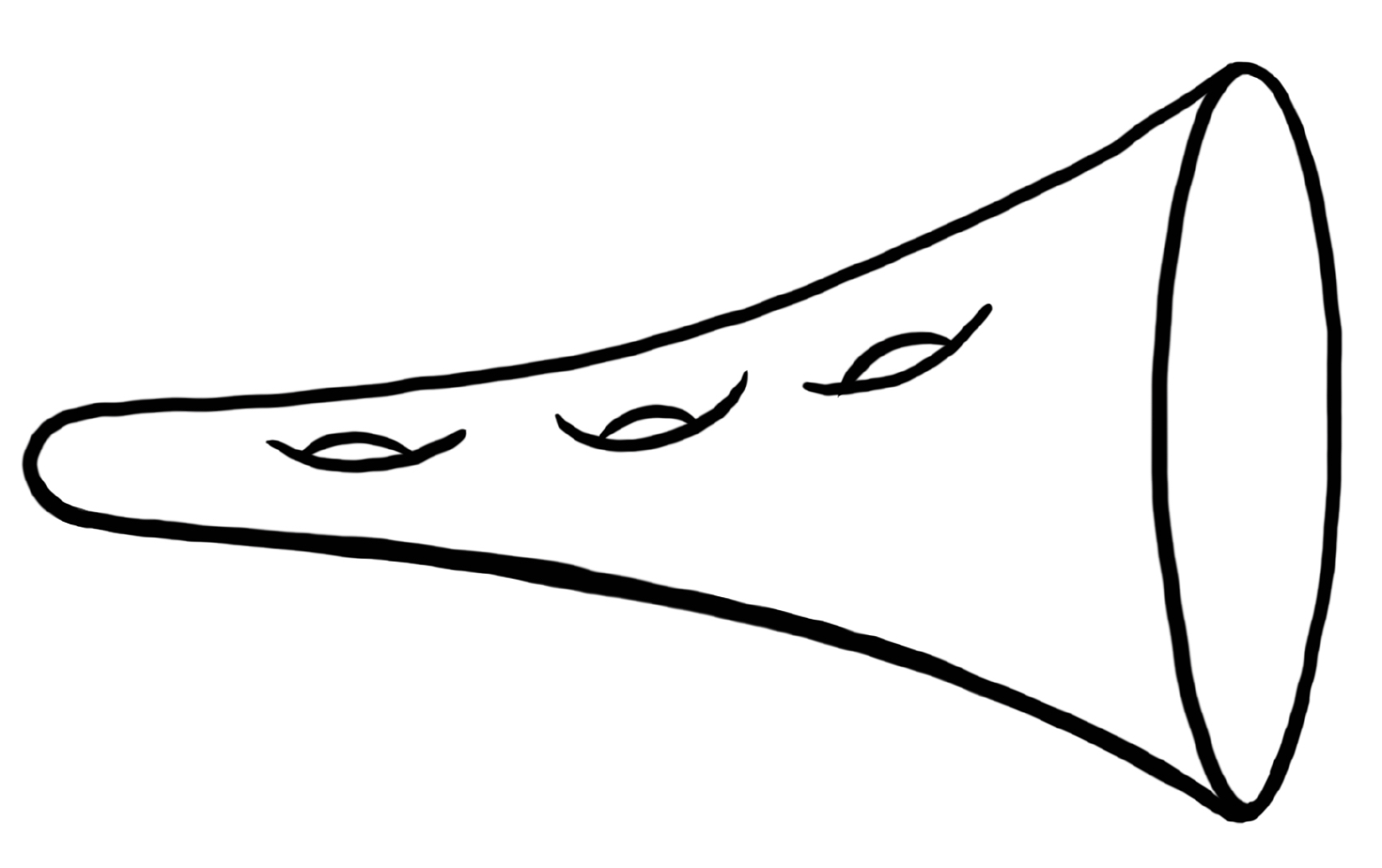}}}

\title{Consistency Conditions for Non-Perturbative Completions of JT Gravity}
\author{Clifford V. Johnson}
\affiliation{\medskip\\ Department of Physics, Jadwin Hall,  Princeton University,
Princeton, NJ 08544-0708, U.S.A.\vskip0.15cm}
 \affiliation{\medskip\\ Department of Physics and Astronomy, University of
Southern California,
 Los Angeles, CA 90089-0484, U.S.A.}


\begin{abstract}
This is a  careful examination of the key components of a large $N$ random matrix model method  for going beyond ordinary JT gravity's topological expansion to define non-perturbative physics. It is offered as  
a simple and (hopefully) clear framework within which {\it any} 
proposed non-perturbative definition should fit, and hence be readily compared to others. Some minimal requirements for constructing consistent non-perturbative formulations are emphasized. A  family of non-perturbative completions emerges from this, which includes an earlier construction. End-of-the-World branes, or simply  D-branes, emerge straightforwardly in this framework and    play a natural  role. 
The many-body fermion picture of the matrix model is a key organizing motif,  with many features highly analogous to a quantum  black hole system, including a  size that grows with  the number of  its microscopic constituents and a  locus (the Fermi surface)  beyond which quantities are traced over in order to define the physics. A formula for the thermal density matrix is proposed that allows a von Neumann form for the entropy to be written in matrix model terms.

\end{abstract}


\maketitle

\section{Introduction}

\label{sec:introduction}

 Jackiw--Teitelboim (JT) gravity~\cite{Jackiw:1984je,Teitelboim:1983ux} and several variants and deformations thereof have been shown (starting with the works of Saad, Stanford and Shenker\cite{Saad:2019lba} and Stanford and Witten~\cite{Stanford:2019vob}) to be perturbatively (in the topological expansion parameter $\hbar\,{\equiv}\, \e^{-S_0}$) equivalent to various double-scaled random matrix models. Here, $S_0$ is the extremal entropy. It is an Hermitian matrix model for ordinary JT gravity, and ref.~\cite{Saad:2019lba}  also pointed out that the model is afflicted by a non-perturbative  instability. A stable definition, needed to address many key physical questions about the gravitational system, had to found.


In ref.~\cite{Johnson:2019eik}, a non-perturbatively stable  definition was presented, and   it was shown~\cite{Johnson:2020exp}  to be  amenable to explicit computation of many key results (such as the full spectral density and the complete spectral form factor). Moreover a  further step was provided recently in ref.~\cite{Johnson:2021zuo} where explicit microstate physics was uncovered non-perturbatively in the form of the probability distributions for the individual energy levels of the underlying ensemble. This was used as an essential ingredient for computing the quenched free energy of the system, allowing the thermodynamics   to be followed all the way down to $T{=}0$.  Such physics is inaccessible without a non-perturbative definition, along with appropriate tools for extracting results.

The definition was presented as based upon double-scaled complex matrix models $M$, appearing in the combination $MM^\dagger$, and hence equivalent to positive Hermitian matrices. This is  regarded by some as a puzzling feature.
It is reasonable to wonder if this non-perturbative completion of JT is unique, or the very least,  natural in some way.  The analogous non-perturbative definitions provided (using the same   methodology~\cite{Johnson:2020heh,Johnson:2020exp,Johnson:2020mwi,Johnson:2021owr}) for various supersymmetric Jackiw--Teitelboim (SJT) models (defined perturbatively as matrix models in ref.~\cite{Stanford:2019vob}) seem to be a much tighter fit, not just because of the positivity, but also because they naturally reproduce (perhaps even explain) several special perturbative features of certain SJT models quite readily. So a concern might be that  ref.~\cite{Johnson:2019eik}'s completion of JT, while of course agreeing with ref.~\cite{Saad:2019lba}'s Hermitian matrix model definition to all orders, is merely an accidental oddity, eventually to be replaced by something more physically compelling.  

This paper will provide several reasons to dismiss such concerns, and  will demonstrate that the non-perturbative completion of ref.~\cite{Johnson:2019eik} is part of a  natural (and tightly constrained) {\it continuous family} of completions that are in fact {\it all Hermitian matrix models}. Moreover, the framework seamlessly incorporates ``End-of-the-world'' branes ({\it i.e.,} D-branes) into JT gravity 
and it will be shown that the families of completions are all dynamically connected, in string theory terms.

Meanwhile, an interesting alternative scheme for {\it potentially} defining JT gravity non-perturbatively, starting with the perturbative Hermitian matrix model tools, was outlined by Saad, Shenker and Stanford~\cite{Saad:2019lba}, and expanded upon recently by Gao, Jafferis,  and Kolchmeyer in ref.~\cite{Gao:2021uro}. It moves off the line of real eigenvalues into the complex plane in order to avoid the (semi-classically established) instability.  However, so far no fully non-perturbative computations for the scheme have been done. It remains an interesting prescription for integration contours, along with  the leading spectral density, and a proposal to use an ensemble of random normal matrices for the definition.   Nevertheless it is reasonable to wonder  if, assuming it can be non-perturbatively explored, it shares any physics in common with the completion of ref.~\cite{Johnson:2019eik}, and its relatives uncovered here. Furthermore, other non-perturbative completions could be proposed in the future ({\it e.g.,} perhaps based on the interesting resurgence work of ref.~\cite{Gregori:2021tvs}), for  which the same questions will arise. 

The broad questions to be answered about all such completions (beyond issues of self-consistency) are: 

\begin{itemize}[label={\raisebox{2pt}{\tiny\textbullet}},leftmargin=*,itemsep=-0.5ex]  
\item{\it Are some more well-motivated (``natural'') than others?}
\item{\it What   physical features   do  they have in common?  }
\item{\it In what ways does the physics differ among them?}
\end{itemize}

%
%

The results and  remarks of this paper will hopefully go some way to clarifying well enough how the  framework (of ref.~\cite{Johnson:2019eik} and this paper) operates in order to provide a precise and clear language for making progress on these questions. The idea is that any proposed non-perturbative completion of JT gravity should in principle correspond to a statement about the behaviour of a very simple function from which stems all the physics. The extent to which completions differ can then be discussed in terms of that function. (The reasoning presented  here for JT and its Hermitian matrix model framework extends  to the SJT cases mentioned above, and any alternative non-perturbative completions that might emerge for those.)


\noindent An outline and summary of the paper's results follows: 

{\bf Section~\ref{sec:toolbox}} begins (in \ref{sec:quantum-mechanics}~and~\ref{sec:meaning}) with a  reminder of  the toolbox of the non-perturbative  framework that emerges from taking a double-scaling limit of a  Hermitian matrix model. 
Some key aspects of the meaning of how the tools are used are emphasized, as a guide to the intuition about perturbative {\it vs} non-perturbative matters.   In particular, there is a known equivalence to a fermionic many-body  system.  A parameter $x\in\mathbb{R}$ parameterizes an excitation energy in that system. The Fermi sea is the region $-\infty\leq x\leq \mu$, where $\mu$ is the Fermi level.  A complete model of perturbation theory is a specification of how the physics is arranged in the Fermi sea.  It comes in the form of an expansion about the leading behaviour of a function~$u(x)$ in the Fermi region, out of which all quantities such as the partition function, and correlation functions thereof, can be constructed.  Completing the model non-perturbatively  must also include a  specification of the behaviour of $u(x)$ in the range $\mu< x\leq+\infty$, which will sometimes  be referred to as the ``trans-Fermi'' region. This cannot be done arbitrarily. The constraints on doing it consistently are a centrepiece  of this paper. 

At this point in the paper a  moderately  attentive reader will perhaps have  noticed similarities between the many-body system with $N$ constituents and a Fermi surface on the one hand, and on the other, a quantum gravity system with $N$ microstates and an horizon. This is emphasized in Section~\ref{sec:Dyson-Gas} since it is a useful organizing structure, and possibly even a dual relationship. It could be useful as a  laboratory  for studying issues of interest in black hole physics,  well beyond the (topological) perturbative regime.
 Pursuing the similarity immediately suggests a formula for a thermal density matrix  for the system, built from tracing out physics on one side of the Fermi level, in complete analogy with a black hole.  The entropy of the system can then be re-written in  von Neumann form, using the trace that refers to the many-body (matrix model) system. A R\'enyi entropy generalization is  straightforward. More properties of the system can be cast into this framework, but since this paper is not specifically about black holes, further elaboration will have to wait for a later publication.\footnote{Nevertheless, it is hard to resist mentioning also that since there are now laboratory experimental systems available that  create many-body Dyson gas-like  systems and apply  tunable potentials, it seems feasible that some aspects of the effective  quantum black hole dynamics could be experimentally accessible.}

{\bf Section~\ref{sec:non-perturbative}} begins the core discussion of the non-perturbative physics. The first part, Section~\ref{sec:semi-classical}, derives and discusses the form of the ``first draft'' of the non-perturbative physics that is to be anticipated based only on perturbative results. It is an alternative  derivation of  formulae given by Saad, Shenker and Stanford~\cite{Saad:2019lba}, and should be thought of as useful semi-classical physics that points the way. The perspective gained using  the methods used here  clearly illustrates the limitations of these results: They need  trans-Fermi data.

Section~\ref{sec:string-equations} starts with a reminder of the core problems with the basic Hermitian matrix model description of JT gravity, as seen in a semi-classical analysis, reviewing some known observations of ref.~\cite{Saad:2019lba} but adding some new observations  about the leading part of $u(x)$ in the trans-Fermi region that hopefully serve to clarify how the framework operates, by  connecting the two approaches. The standard matrix model supplies  a non-linear differential equation~(\ref{eq:simple-string-equation})  for $u(x)$, which for historical reasons is called the ``string equation''.  Non-perturbative problems translate directly into its failure to extend $u(x)$ properly into the trans-Fermi region.   
The argument is made that, whatever its origin,  {\it any non-perturbative completion} of the perturbative matrix model  physics should be equivalent to specifying the full $u(x)$, including within the  trans-Fermi region. In this way, different completions can be quantitatively compared.  

A basic consistency condition on non-perturbative extensions is outlined, in terms of how $u(x)$ can behave as it extends into the trans-Fermi regime, even at leading order (extending observations made in ref.~\cite{Johnson:2020lns}). The identification of the correct string equation to use, equation~(\ref{eq:big-string-equation}), is discussed.  It is also directly derivable from a random Hermitian matrix model~\cite{Dalley:1991xx,Dalley:1992yi}, which lends support to the naturalness of the completion in this context. The prototype non-perturbative completion of ref.~\cite{Johnson:2019eik} is based on a special case, $\sigma{=}0$ and $\Gamma{=}0$, of this equation.\footnote{\label{foot:remark}A longer remark is in order here: The  equation~(\ref{eq:big-string-equation}) with $\sigma{=}0$ was first derived in the context of complex matrix models $M$, but with a potential built from $MM^\dagger$, and so can be thought of as a model of positive Hermitian matrices. It was used in refs.~\cite{Johnson:2019eik,Johnson:2020heh,Johnson:2020exp,Johnson:2020mwi} (in two different ways)  to non-perturbatively define both JT gravity and certain models of super JT gravity. In the latter case the positivity constraint is natural, since $H=Q^2\geq0$, but in the former case, while   not wrong, it seems unneccessary. The resolution is that it is better thought of, in the bosonic JT case, as the  $\sigma{=}0$ case of an equation 
for random Hermitian matrices restricted to have lower bound $\sigma$ on their spectrum.}

{\bf Section~\ref{sec:consistency-conditions}} begins by laying out a tentative set of extra  consistency conditions, and it is immediately apparent that there is a family of distinct non-perturbative completions that satisfy them. They are described, and allow for new solutions of equation~(\ref{eq:big-string-equation}), with $\sigma$ now taking some non-zero value that characterizes the completion. These consistency conditions constrain the allowed values of $\sigma$. The lowest possible value  in these simple cases (without background D-branes) is $-(j_{01}/2\pi)^2{\simeq}-0.14648...$, a  result that is considerably higher than the $E{=}{-}\frac{1}{4}$  value suggested  semi-classically. The possibility of lower values of $\sigma$ that evade the conditions above is discussed. Section~\ref{sec:explicit} computes and exhibits examples (for various values of $\sigma$) of the basic function $u(x)$, and the resulting  non-perturbative spectral densities, two-point correlation functions, and microstate spectra, discussing the similarities and differences among the various completions. For small $\hbar$ the differences amount to an instanton effect controlled by~$\sigma$.

{\bf Section~\ref{sec:D-branes}} shows that it is natural to enhance these models with what are commonly called End-of-the-World branes in this context~\cite{Penington:2019kki}. They are simply a kind of background D-brane that is  known~\cite{Dalley:1992br} to be described by simply turning on $\Gamma$  (which counts their number) in the string equation~(\ref{eq:big-string-equation}). Perturbatively it is equivalent to tuning the underlying KdV (closed string) parameters~$t_k$  in such a way as to turn on open string sectors, a transformation noted in this context  long ago in   refs.\cite{Itoh:1992np,Johnson:1994vk}. They are incorporated into the consistency conditions and their effects  described with the aid of a useful 't Hooft limit  of taking $\Gamma{\to}\infty$  and $\hbar{\to}0$ holding   $\hbar\Gamma$  finite. The string equation becomes an algebraic constraint in this limit, where the key features can be readily extracted.  The overall picture is that, if desired, background D-branes can be used to enlarge the family of non-perturbative completions because they act to add stability to the overall setup by repelling eigenvalues away from the unstable zone.  A simple sign change, making the branes attractive instead of repulsive, has the opposite effect.

In {\bf Section~\ref{sec:dynamical-boundaries}}, string theory language and intuition, lurking throughout this framework, is unveiled to make it clear that all the features seen have a very natural organization. This is all just the physics of the dynamical interplay of open and closed string sectors. The key non-perturbative parameter,  $\sigma$  simply makes a non-perturbative contribution to the boundary cosmological constant (the coefficient of a  closed string operator). The background D-branes (if turned on) incorporate it into their world volume. 
Deformations of JT gravity will also dynamically induce changes of $\sigma$, and these are described infinitessimally by a family of modified Virasoro constraints that naturally act on the $\tau$-function defined {\it via}:  $u(x,t_k;\sigma){=}{-}\hbar^2\partial^2_x \ln\tau$. Virasoro contraints are the manifestation of the diffeomorphism invariance of the spectrum, which naively runs over the whole line for Hermitian matrix models. Perturbatively, $\sigma$ is invisible, and the string equation to use is the $L_{-1}$ constraint, as usual. Non-perturbatively, from the consistency conditions seen here, the spectrum {\it must} end at some $\sigma$, and so $L_{-1}$ (translations of $\sigma$) is broken, leaving $L_0$ (scaling) as the string equation, which is precisely equation~(\ref{eq:big-string-equation}).
It is also remarked in this section that  since this string equation follows from assuming just a scaling symmetry and the  KdV-flow structure that is present perturbatively in the Hermitian matrix models,  non-perturbative completions can be further classified as to whether they violate these assumptions, or not, in which case they must supply a $u(x)$ that satisfies the string equation.

{\bf Section~\ref{sec:discussion}} contains some concluding thoughts.  

\section{The Toolbox}
\label{sec:toolbox}
The key tools needed are very simple to describe. However, they are used in a way that can seem counterintuitive. An attempt will be made in the next few subsections to explain the intuition behind the various quantities and expressions to follow, with an eye on emphasizing features and new insights  that help inform the non-perturbative methodology on which  this paper focuses.\footnote{A forthcoming paper~\cite{metoappear}, in an even longer introduction, will further unpack and explain how the toolbox works, with several new insights.} 
\subsection{A Model of Quantum Mechanics }
\label{sec:quantum-mechanics}
The main workhorse is a quantum mechanics problem with Schr\"odinger operator:
\be
\label{eq:schrodinger}
{\mathcal H} = -\hbar^2\frac{\partial^2}{\partial x^2}+u(x)\ ,
\ee 
($u(x)$ will be defined shortly) defining a family of wavefunctions $\psi(E,x)$ {\it via} the spectral problem
\be
\label{eq:spectral}
{\cal H}\psi(E,x)=E\psi(E,x)\ .
\ee
The partition function of  JT gravity system is then:
\be
\label{eq:JTpartfun}
Z(\beta) = \int_{-\infty}^{\mu}\langle x|\e^{-\beta{\cal H}}|x\rangle dx\ ,
\ee
where $\mu$ will be discussed shortly. The  topological counting parameter, $\hbar$, (renormalized $1/N$ from the matrix model perspective) is related to the extremal entropy of JT gravity {\it via} $\hbar{\equiv}\e^{-S_0}$. From~(\ref{eq:JTpartfun}) follows (insert a complete set of energy eigenstates) the spectral density (the Laplace transform of $Z(\beta)$):
\be
\label{eq:spectral-density}
\rho(E) = \int_{-\infty}^{\mu} |\psi(E,x)|^2 dx\ .
\ee
The appearance of this quantum mechanical system, and how it is used ({\it e.g.} the equation just written), is often regarded as mysterious, but its origin and role are very simple to state. It is best understood in the context of 
the matrix model's elegant organization as a fermionic many-body system.\footnote{This picture goes all the way back to work such as refs.~\cite{Bessis:1980ss}, and was developed into a powerful tool in the 2D gravity context in refs.~\cite{Gross:1990aw,Banks:1990df}.}  This will be discussed below in Section~\ref{sec:meaning}, after a pause to connect to perturbation theory.

Perturbation theory about some leading order solution (a large $N$ saddle of the matrix model) is equivalent to knowing the function $u(x)$ in the $-\infty\leq x\leq\mu$ region as an expansion in $\hbar$ about some leading solution $u_0(x)=\lim_{\hbar\to0} u(x)$:
\be
\label{eq:potential-perturbative}
u(x)=u_0(x)+\sum_{g=1}^\infty u_g(x)\hbar^{2g}+\cdots\ ,
\ee
and the ellipsis denotes non-perturbative contributions to be discussed later. Note that, at any order that $u(x)$ is known to, the quantum mechanics problem~(\ref{eq:schrodinger}) using it as a potential  can itself have its own $\hbar$ expansion. 
 For small $\hbar$ (or large $E$)  inserting the leading WKB form of the wavefunctions $\psi(E,x)$ turns equation~(\ref{eq:spectral-density}) into, at leading order (using just $u_0(x)$ as the potential):
\be
\label{eq:leading-relation}
\rho_0(E)=\frac{1}{2\pi\hbar}\int_{-x_{\rm c}}^\mu\frac{dx}{\sqrt{E-u_0(x)}}\ ,
\ee
where $-x_{\rm c}$ is where the square root vanishes.  A convenient normalization for $\psi(E,x)$  has been chosen here to match to perturbation theory. This includes a $1/\sqrt{\hbar}$ factor in the wavefunction that yields a leading density that scales with $N$, which fits nicely with $\hbar$ being the (renormalized) $1/N$ of the matrix model, the topological expansion parameter {\it \`a la} 't Hooft~\cite{'tHooft:1973jz}. (The explicit WKB form, is given later in equation~(\ref{eq:WKB}) where it will be further discussed.)
The defining equation for $u_0(x)$ in this limit is:
\be
\label{eq:leading-R0}
{\cal R}_0\equiv\sum_{k=1}^\infty t_ku_0^k+x=0\ ,
\ee
where this form  incorporates all the possible ``multicritical'' behaviour (in the sense of ref.~\cite{Kazakov:1989bc}).
Here, parameter~$t_k$ brings in an admixture of  the behaviour associated with the $k$th ``minimal model''.
 Matching to JT gravity's (Schwarzian) partition function~\cite{Jensen:2016pah,Maldacena:2016upp,Engelsoy:2016xyb} means that  the spectral density should be:
 \be
 \label{eq:leading-density}
 \rho_0(E)=\frac{\sinh(2\pi\sqrt{E})}{4\pi^2\hbar}
 \ee
which sets the $t_k$ and $\mu$  uniquely to~\cite{Dijkgraaf:2018vnm,Johnson:2019eik}:\footnote{It is worth remarking here that there are several things called minimal models, and $t_k$, in the literature, and the basis of operators they couple to. The approach used here is to use what might be called  the KdV minimal models, with the $t_k$ that appear most naturally in the matrix model: The model controlled by $t_k$ simply yields a behaviour $E^{k-\frac12}$ in the spectral density.  Even within the KdV convention, there are different common normalizations for the $t_k$ arising from whether the coefficient of~$u^k$ in~(\ref{eq:leading-R0}) is chosen as unity or not. Section~\ref{sec:dynamical-boundaries} will discuss this further. The KdV basis is  different from the operator basis more naturally occurring in the conformal minimal modelsl~\cite{Moore:1991ir}, and so the $t_k$s in such models are a mixture of the $t_k$s  used here, as nicely laid out in this context in ref.~\cite{Mertens:2020hbs}. After translating, the approach used here  aligns nicely with the large $k$  approach first suggested in ref.~\cite{Saad:2019lba}. CVJ thanks Joaquin Turiaci for a comment about this.}
\be 
\label{eq:teekay}
t_k{=}\frac{\pi^{2k-2}}{2k!(k-1)!}\ ,\quad \mu=0 \ ,
\ee  giving:
 \be
 \label{eq:leading-u}
 {\cal R}_0\equiv\frac{\sqrt{u_0}}{2\pi}I_1(2\pi\sqrt{u_0})+x =0\ .
 \ee
 The parameter $\mu$,  fixed to zero here, has the interpretation as  a bulk gravity cosmological constant in the $k{=}2$ model. For more general $k$ it is the coefficient of the lowest dimension closed string or ``bulk'' operator, and matching to perturbation theory shows  that it is set to zero. The next Section will emphasize that it also has an interpretation as a Fermi level in the fermionic many-body description of the matrix model, and the entire region $-\infty\leq x\leq \mu$ pertains to  the Fermi sea making up the many-body state. (Although $\mu{=}0$ for JT, it will often be still referred to as $\mu$ in several expressions to follow, as a reminder of its Fermi level role.)

For the matrix model, perturbation theory is fully  described in the $x{<}\mu$ region. Corrections to $u_0(x)$   in this region can be developed as an asymptotic series of the form~(\ref{eq:potential-perturbative}), determined by an ordinary differential equation called a ``string equation'' which will be discussed in Section~\ref{sec:string-equations}.   Since $x$-derivatives will come with a factor of $\hbar$ in the equation, the  effective expansion parameter of the equation is in fact~$\hbar/|x|$,  so for a given value of $\hbar$, the asymptotic series is also an  expansion for $u(x)$ about large $|x|$, deep in the Fermi sea. 

Understanding of physics beyond perturbation theory will ultimately come from knowledge pertaining to  what shall be called  the ``trans-Fermi'' region $\mu< x\leq+\infty$. This should be clear from the statement of the physics in terms of a  quantum mechanics problem on the whole of $x$, over which the wavefunctions can spread, and contribute to the integral~(\ref{eq:JTpartfun}) even in classically forbidden regions. There are two types of non-pertubative correction in fact: The first  are  non-perturbative  quantum mechanical results 
that come even if $u(x)$ is only known perturbatively, following from the  form of  wavefunctions (which exist on the whole $x$ line).   These are discussed in Section~\ref{sec:semi-classical}. The second type are non-perturbative contributions to  $u(x)$, which need to be understood in order to make sense of the whole theory. While such corrections will manifest themselves in the Fermi sea (the large $|x|$ asymptotic expansion of $u(x)$ will eventually break down as $|x|$ becomes small, well before $x=\mu$), their full understanding is in terms of how $u(x)$ extends as a smooth function over the whole real $x$ line. As already mentioned, this  is   governed by a string equation, and will be  discussed in Section~\ref{sec:string-equations} and even further in Section~\ref{sec:consistency-conditions}.

\subsection{The Meaning of the Quantum Mechanics}
\label{sec:meaning}
\begin{figure}[h]
\centering
\includegraphics[width=0.45\textwidth]{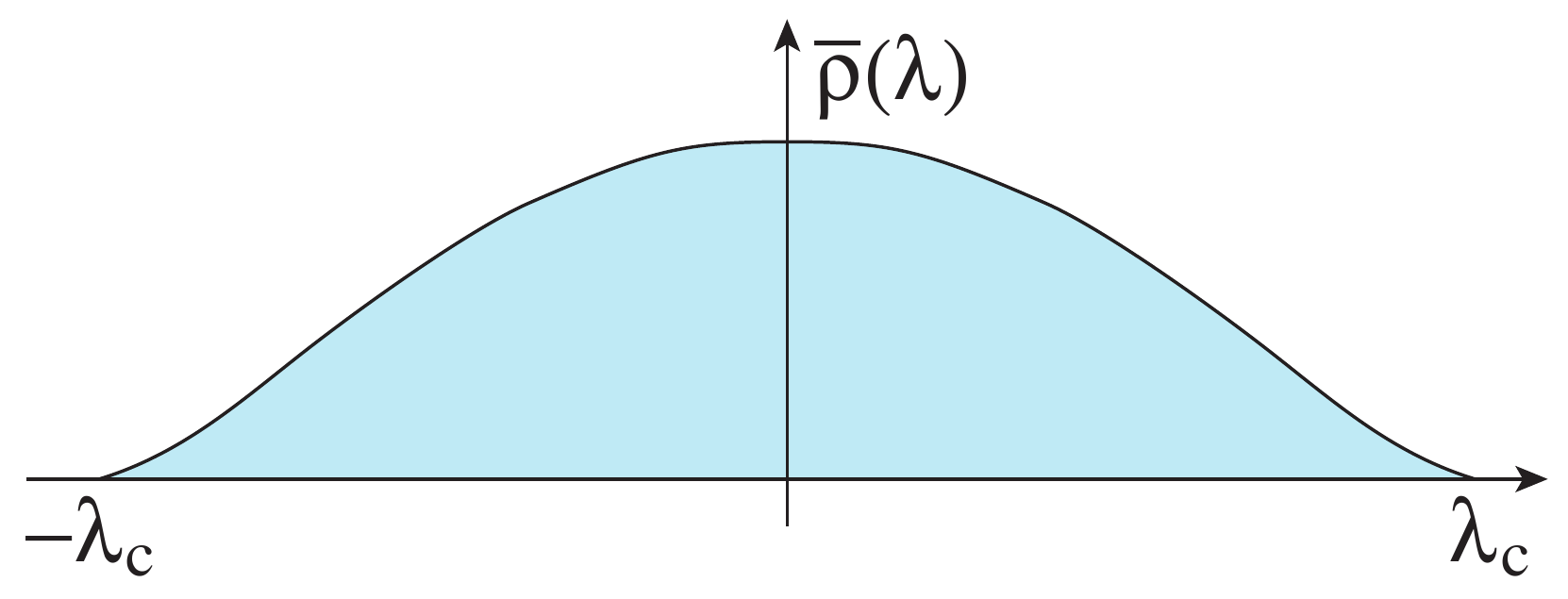}
\caption{\label{fig:dyson-gas} The Dyson gas at large $N$.}
\end{figure}

Recall that the $N\times N$ matrices can be diagonalized to the $N$ eigenvalues $\lambda_i$ ($i=1,\cdots,N$), and the matrix model is then a system of particles on a line with positions $\lambda_i$, subject to  a potential which is a combination of the original potential $V(\lambda_i)$ of the matrix model supplemented by a repulsion term (coming from the (squared) Vandermonde determinant Jacobian in going from~$M$ to~$\lambda_i$). This is the Dyson gas description.\footnote{There are several excellent reviews of this. See {\it e.g.} refs.~\cite{Meh2004,ForresterBook,Eynard:2015aea}.} The leading large $N$ description is as a ``droplet'' saddle point solution with ends (for even $V(\lambda)$) at $\pm\lambda_c$ with shape given by the density function ${\bar \rho}(\lambda)$, where $\lambda$ is a continuous coordinate in this limit. See figure~\ref{fig:dyson-gas}.

 A key point here is that the precise description of the Dyson gas dictated by the matrix model is equivalent to a many-body fermionic system. Each $\lambda_i$, in potential $V(\lambda_i)$ can be described as being in an excited state, with a wavefunction that can be written as ${\tilde \psi}_n(\lambda_i){=}\e^{V(\lambda_i)/2} P_n(\lambda_i)/\sqrt{h_n}$, where   $P_n(\lambda_i)\sim\lambda_i^n+\cdots$ ($n\in\mathbb{Z}^+$) are a family of orthogonal polynomials.\footnote{Their orthogonality is with respect to the matrix model potential:  $\int P_n(\lambda)P_m(\lambda) \e^{-V(\lambda)}d\lambda{=}h_n\delta_{nm}$.}  The integer $n$ correlates with the energy of the excitation. ({\it c.f.} the  quantum harmonic oscillator, where these wavefunctions are simply the Hermite functions.)  The free fermion system intrinsic to the matrix model is simply   built by exciting a particular pattern of oscillations across all the $\lambda_i$  in such a way as to build a many body state $|\Psi\rangle$ which is a Slater determinant.  It is not a choice, but  is simply a reflection of the appearance of the squared Vandermonde determinant in the Dyson gas picture, which produced the logarithmic repulsion that stops the $\lambda_i$s from coinciding.    The first $N$ of the orthogonal polynomials are used in building the model in this way, and this is equivalent to filling up the ``Fermi sea''  of energy levels $n=0,1,\cdots,N$. 

At large~$N$, a smooth coordinate $X {=} n/N$ can be used, running from $0$ to 1 in going from the sea floor to the surface, the Fermi level at $X=1$.  Various results of computing a physical quantity  in the matrix model, when evaluated in terms of orthogonal polynomials, can be written as sums of quantities over all the $N$ levels, and at large~$N$ these become integrals over $X$ from 0 to~1.  Crucially, {\it the double-scaling limit focuses on a scaled neighbourhood of the Fermi level:} Writing $X{=}1{+}(x{-}\mu)\delta^{2k}$ (where~$k$ is some positive power that indexes the $k$th minimal model), and where $\delta\to0$ as $N\to\infty$, the parameter $x$  runs from $-\infty$ to $\mu$ in spanning what is now a scaled Fermi sea, with Fermi level at $x=\mu$. So now it is clear that coordinate $x$ in the quantum mechanics~(\ref{eq:schrodinger}) is actually parameterizing an {\it energy} in the matrix model, and the integrals in the equations above are instructions for how to build certain objects in the many body system, filling up to the Fermi level.  Energy in the quantum mechanics is instead, amusingly, position in the Dyson gas or many-body system.  Table~\ref{tab:dictionary1} summarizes a few of these dictionary entries for easy reference. 
\begin{table}[h]
\begin{center}
\begin{tabular}{|l|l|}
\hline
\textbf{Quantum} &\textbf{Dyson Gas or}\\
\textbf{Mechanics} &\textbf{Many-Body System}\\
\hline\hline
\textrm{Energy $E$}&\textrm{Position $\lambda_i\to\lambda=\lambda_{\rm c}-E\delta^2$}\\
\hline
\textrm{Position $x$}&\textrm{Energy level $\frac{n}{N}\to X=1+(x-\mu)\delta^{2k}$}\\
\hline
\textrm{       $-\infty\leq x\leq\mu$}&\textrm{The Fermi sea.}\\
\hline
\textrm{ $x=\mu$}&\textrm{Fermi level $X=1$}\\
\hline
\textrm{       $\mu<x\leq+\infty$}&\textrm{The trans-Fermi regime.}\\
\hline
\textrm{Planck's $\hbar$}&\textrm{ $\frac{1}{N}_{\phantom{1}}=\hbar\delta^{2k+1}$}\\
\hline
\textrm{Wavefunction: }&\textrm{Excited state $n$ at position $\lambda_i$:}\\
\textrm{       $\psi(x,E)$}&\textrm{     ${\tilde \psi}_n(\lambda_i)=\e^{-\frac{V(\lambda_i)}{2}}P_n(\lambda_i)/\sqrt{h_n}$.}\\
\hline
\end{tabular}
\end{center}
\caption{Part of  a dictionary translating Quantum Mechanics quantities to those in the Matrix Model/Dyson Gas/many-body system. Here $k$ refers to the $k$th critical/minimal  model.}
\label{tab:dictionary1}
\end{table}

The detailed origin of each piece of the toolbox is not needed here, and moreover there are many other useful pieces that won't be used in this paper. But there is a very simple and clear  dictionary translation between each piece of the toolbox and an aspect of the many-body or Dyson gas system and hence the matrix model.  

The fundamental tool for extracting the answers to many physics questions is primarily the wavefunction~$\psi(E,x)$, which is the double-scaled limit of ${\tilde \psi}_n(\lambda_i)$, (which recall is the $n$th oscillator state at position $\lambda_i$). This can  be seen for example for the spectral density~(\ref{eq:spectral-density}), and hence the partition function $Z(\beta)$, but  it is also the basic object to use for computing  correlators of arbitrary numbers of copies of $Z(\beta)$, and more besides.\footnote{\label{fn:kernel} More succinctly, $\psi(E,x)$ can  be used to construct the matrix model kernel \bea K(E,E^\prime)&=&\int_{-\infty}^\mu \psi(E,x) \psi(E^\prime,x) dx \nonumber\\
&=& \frac{\psi(E,\mu)\psi^\prime(E^\prime,\mu)-\psi(E^\prime,\mu)\psi^\prime(E,\mu)}{E-E^\prime}\nonumber\ ,\eea from which many quantities can be computed. The last line follows from a Christoffel-Darboux identity, which usefully writes everything in terms of the wavefunction and its $E$-derivative   at the Fermi surface. Note that $\rho(E){=}K(E,E)$.}  So the role of the quantum mechanics is to define $\psi(E,x)$, and this is done through knowing the potential, $u(x)$.  Its origin in the matrix model is also straightforward. It is the double-scaling limit of the recursion coefficient that determines an orthogonal polynomial at a given order in terms of polynomials at lower order: {\it i.e.,} $\lambda P_n=P_{n+1}+R_nP_{n-1}$, for even $V(M)$. The $R_n$, (which becomes $R(X)$ at large~$N$) are ultimately determined from the matrix model potential $V(M)$, and so $u(x)$ (the scaled piece of $R(X)$) is in a sense the embodiment of $V(M)$ once the double-scaling dust has settled. As already stated, the physics is built from the $P_n$ (and how they are assigned among the $\lambda_i$). A most natural way that a complete set of $\psi(E,x)$ can be determined given some basic function $u(x)$ is to have them be a complete set of wavefunctions in some potential $u(x)$. That is precisely in accord with  the appearance of the Hamiltonian ${\cal H}$, equation~(\ref{eq:schrodinger}). It is in fact the double-scaling limit of the operator for multiplying  the $P_n$ by $\lambda$: The quantum mechanics' energy operator becomes  the position operator in the Dyson gas.

So knowing the potential $u(x)$ determines  the set of wavefunctions $\psi(E,x)$, and hence specifies the whole  model. Put differently, $u(x)$ contains the DNA of what the entire many-body system is built from, for a particular matrix model, and hence the gravity system. The Schr\"odinger equation~(\ref{eq:schrodinger}) then expresses this DNA into $\psi(E,x)$, from which JT gravity quantities are built by inserting them into the many-body expressions, equations~(\ref{eq:JTpartfun}) and~(\ref{eq:spectral-density}). 

Section~\ref{sec:string-equations}~and~\ref{sec:consistency-conditions} will  discuss the equations governing the complete form of $u(x)$, historically known as ``string equations'' allowing for the complete extraction of the complete non-perturbative physics of the model.

\subsection{Another View of the Dyson Gas}
\label{sec:Dyson-Gas}
Before proceeding with discussions of non-perturbative physics, it is worth making some observations about an alignment between the properties of the matrix model (described as a Dyson gas and as a many-body system) and those of quantum gravitational systems, especially black holes. They are not offered as amusing accidents, but as dual phenomena perhaps  key to understanding why the matrix models can capture aspects of black holes so well. 

The Dyson gas is made from $N$ constituents that collectively interact repulsively to puff the system up to a size that grows with $N$. (Usually, in preparation for large~$N$, a rescaling is performed that obscures this growth, but it is there.) A key feature of quantum gravity is an analogous growth, manifested in the fact that a black hole's horizon size is set by the number of microstate constituents. This is spelled out in the Bekenstein-Hawking formula\cite{Bekenstein:1973ur,Bekenstein:1974ax,Hawking:1974sw,Hawking:1976de} linking horizon area to  entropy: 
${A/4}=S=\log N$.

The next observation is about the fermionic many-body description of the Dyson gas, which defines a coordinate $x$ and a Fermi-surface at $x=\mu$. All physical properties of the system, from the spectrum and spectral density $\rho(E)$ through to all correlators, involve summing over all the dependence within the Fermi sea $(-\infty\leq x\leq\mu)$. The full quantum mechanics on $x$ has Hamiltonian ${\cal H}$ but the Dyson gas is described by   projecting or tracing out the physics  on $x$ using ${\rm Tr}_{\cal P}[\bullet]{\equiv}\int_{-\infty}^\mu \langle x|\bullet|x\rangle dx$ which combines summing over the Hilbert space with projecting out the unwanted part. This builds the many-body expressions described earlier. The symbol ${\cal P}$ is a reminder of the projection present: Only integrate up to the Fermi surface. This is very reminiscent of how thermal properties of a black hole (and horizons, more generally) are often described in quantum gravity too. Tracing out one side results in a thermal vacuum, nicely described in terms of a thermal (Gibbs) density of state. 

It would be nice to bring the two descriptions even closer, showing that, for example, the entropy derived from the Dyson gas (and hence the JT gravity physics it captures) has a von-Neumann description.  It's not hard to guess what to do (at least for high temperatures where the distinction between quenched and annealed formulations can be safely ignored) once it is recalled that the matrix model partition function~(\ref{eq:JTpartfun}) is
$Z(\beta){=}\int \rho(E)\e^{-\beta E}dE={\rm Tr}_{\cal P}[\e^{-\beta{\cal H}}]$. The matrix model thermal density matrix would seem to be naturally:
\be
\label{eq:density-matrix}
{\widehat\rho}=\frac{\e^{-\beta{\cal H}}}{Z(\beta)}\ ,
\ee
where a hat is used here to avoid disastrous confusion with the spectral density.
This construction ensures conservation of probability, {\it i.e.,}  ${\rm Tr}_{\cal P}[{\hat\rho}]=1$. Many useful quantities should then be constructible from this density matrix using the trace. So the entropy is:
\bea
\label{eq:von-neumann}
S(\beta) &=& -{\rm Tr}_{\cal P}[ {\widehat\rho\log{\widehat\rho}}] \nonumber\\
&=&  \frac{1}{Z(\beta)}{\rm Tr}_{\cal P} \left[\beta{\cal H}\e^{-\beta{\cal H}}+\e^{-\beta{\cal H}}\log Z(\beta)\right]\nonumber\\
&=&\beta U-\beta F\ ,
\eea
where 
\bea U=\frac{1}{Z(\beta)} \int E\rho(E)\e^{-\beta E}dE
\eea
 and 
 \bea 
 F=-\beta^{-1}\log Z(\beta)\  
 \eea   are the overall energy and  the Helmholtz free energy (annealed) recovering the correct thermodynamic relation, entirely in matrix model terms.  R\'enyi entropies can then be readily be defined  the usual way:
 \be
 S^{(n)}(\beta) = \frac{1}{1-n} \log\left[{\rm Tr}_{\cal P}( {\widehat\rho}^n)\right] \ .
 \ee
 The success of this structure suggests that there is considerable mileage to be obtained in regarding the Fermi surface as acting akin to an entanglement horizon, yielding a rich  thermal character to the physics of an otherwise threadbare  simple quantum mechanics on $x$. Care must be exercised in thinking of $x$ as literally a spatial coordinate (and $x{=}\mu$ as an horizon), since this thermal-like construction is robustly present for a wider variety of models than just ones known to be connected to JT gravity and black holes. Even the  Airy model (the $k=1$ simple Hermitian case), has this thermodynamics, along with $x$ and the Fermi surface. It does not have the critical structure to support finite area surfaces and is hence a topological theory, not thought (yet) to arise as part of a near-horizon black hole story in its own right.

Notice that the free energy arising above in equation~(\ref{eq:von-neumann}) is really the annealed quantity, since $Z(\beta)$ arises by definition as a random matrix average, perhaps better denoted $\langle Z(\beta)\rangle$, and then the logarithm acts on it.  Given what was learned   in ref.~\cite{Johnson:2021zuo} about how to extract the full {\it quenched} free energy $F_Q{=}-\beta^{-1}\langle\log Z(\beta)\rangle$  from non-perturbative information about the matrix model, allowing low temperature physics to be extracted, it might be possible to take this a step further and build a direct definition of $S(\beta)$ in terms of a modified density matrix that involves $F_Q(\beta)$ instead of $F_A(\beta)$. In effect, the difference between annealed and quenched is simply a matter of  taking into account the underlying  statistics that is actually present in the model (extractible from the  kernel  $K(E,E^\prime)$ (see footnote~\ref{fn:kernel} on page \pageref{fn:kernel}) using Fredholm determinants) or coarse-graining over it and using just its diagonal $\rho(E)$. So  a quench-aware thermal density matrix, if it exists, must  use information contained in the more refined object, $K(E,E^\prime)$.  Pursuing this  is not really within the scope of this paper, so will be left for another venue.

\section{Non-Perturbative Physics}
\label{sec:non-perturbative}
With the toolbox and overall framework recalled and contextualized, the central discussion of  non-perturbative matters can begin in earnest. To repeat, all the physics is encapsulated in knowing $u(x)$. As will be discussed shortly, the matrix model provides an ordinary differential equation in $x$ that determines it, which arises as the double scaling limit of a difference equation for orthogonal polynomial recursion coefficients. That equation can be expanded  in $\hbar/|x|$ in the Fermi sea regime $-\infty{\leq}x{\leq}\mu$, to develop a perturbative expansion for~$u(x)$. Quantum mechanics on such a perturbative expansion of $u(x)$ is precisely equivalent to the topological recursion~\cite{Mirzakhani:2006fta,Eynard:2007fi} approach to the matrix model used by Saad, Shenker and Stanford~\cite{Saad:2019lba}. Fully non-perturbative information about~$u(x)$ is needed to complete the story and it requires information about the  trans-Fermi regime $\mu{<} x{\leq}{+}\infty$, and that is what this paper is primarily about. 

The  reader might be puzzled though, because ref.~\cite{Saad:2019lba} seemed to provide some non-perturbative insights into the JT gravity matrix model results without knowing anything more than the  topological expansion. This is what might be called semi-classical non-perturbative information, and it is to be regarded as a useful {\it first draft} of the full non-perturbative physics. It is worth pausing to describe how that works 
before then moving on to the full non-perturbative physics in subsection~\ref{sec:string-equations}.

\subsection{Non-Perturbative Data I: Semi-Classical}
\label{sec:semi-classical}
Regardless of to what order in $\hbar$-perturbation theory $u(x)$ is known  (from expanding the string equation to be discussed in Section~\ref{sec:string-equations}), there are non-perturbative effects that can be readily identified by reference to  the quantum mechanical form~(\ref{eq:spectral-density}) of the expression for $\rho(E)$. This is the case even for the leading (``classical'') disc result $u_0$, for which the leading spectral density is $\rho_0(E)$. There will be non-perturbative contributions of two different kinds, those associated with $E{>}0$ states, and those with $E{<}0$ states. It all follows from the WKB form of the wavefunction, mentioned earlier, and now displayed here:
\bea
\label{eq:WKB}
&&\psi(E,x)\simeq\frac{1}{\sqrt{\pi\hbar}}\frac{1}{[E-u_0(x)]^\frac14} \\
&&\hskip3cm\times\cos\left({\frac{1}{\hbar}\int^x\!\!\!\sqrt{E-u_0(x')}\,dx'}-\frac{\pi}{4}\right)\ , \nonumber
\eea
where a phase (to get cosine from two arbitrary complex exponentials) and an overall  normalization have been fixed.\footnote{This can be done by connecting with the exactly known~\cite{Moore:1990cn} wavefunctions $\psi(E,x){=}\hbar^{-\frac23}{\rm Ai}[-(E{+}x)\hbar^{-\frac23}]$ of the  Airy model ($k{=}1$ simplest Hermitian matrix model), where $u(x){=}u_0(x){=}{-}x$.} Some caution should be exercised here: For a given value of $\hbar$, while this expression improves its accuracy for larger $E$,  for smaller $E$ it will eventually become inaccurate, requiring knowledge about how the potential is corrected near the Fermi surface, and also how the wavefunction (and hence the potential) extends into the trans-Fermi regime. 

In the classical $E>u_0(x)$ region, the dominant contribution is from the pre-factor, as already mentioned, and it yields the integral formula~(\ref{eq:leading-relation}) for the  leading part of $\rho(x)$ that was used earlier. There are non-perturbative oscillatory modulations that have been ignored so far (except for a factor of $\frac12$ they produced in~(\ref{eq:leading-relation}) when averaging over them). They come from the integral over the (squared) oscillating part of the wavefunction, no longer assuming that they will average out: This was fine to assume at high energy, but their frequency decreases with decreasing $E$.
Computing the integral\footnote{A direct approach, mentioned  by Felipe Rosso, gets the form of the second term by using an identity derivable using multiple $\mu$-differentiation.  The approach used here starts with footnote~\ref{fn:kernel}'s  form of the kernel $K(E,E^\prime)$ for which $\rho(E){=}K(E,E)$ (as was done in ref.~\cite{Johnson:2021owr} for an SJT model). Writing $E^\prime{=}E+\epsilon$ and expanding in $\epsilon$ yields the result as $\epsilon{\to}0$.} gives the non-perturbative oscillations at the next order as:
\be
\label{eq:semi-classical-np-1}
\rho_{\rm sc}=\rhoo(E) - \frac{1}{4\pi E}\cos\left(2\pi\!\int^E_0\!\!\!\rhoo(E^\prime)dE^\prime\right)\ , \quad E>0\ ,
\ee
where $\rhoo(E)$ is in equation~(\ref{eq:leading-relation}) and the rewriting 
\be
\label{eq:integrate-momentum}
\int^\mu_{-\infty}\!\!\sqrt{E-u_0(x)}dx = \pi\hbar\!\int^E_0\!\!\!\rhoo(E^\prime)dE^\prime
\ee
was used.
There are {\it also} non-perturbative corrections coming from the fact that there are exponential  tails of the wavefunctions that can penetrate into the region where $E<u_0(x)$, but those will be small compared to the oscillations. 

On the other hand, there are also such exponential tails contributing to the density integral from the $E<0$ sector. The point is that the quantum mechanics is defined beyond $x=\mu$ into the trans-Fermi-regime too, and so the wavefunctions whose osciallatory part is entirely in that regime will still make contributions in the Fermi sea (where the $x$ integral is performed) {\it via} their ``forbidden region" exponential tails. This is another sign that knowledge of $u(x)$ in the trans-Fermi regime is crucial. The best guess as to what the non-perturbative contributions might look like is to assume the same WKB form as above but now with a decaying exponential instead of the oscillatory piece. This throws away half the contributions to the calculation done above and leaves
\bea
\label{eq:semi-classical-np-2}
\rho_{\rm sc}&=& - \frac{1}{8\pi E}\exp\left(-2\pi\!\int^{-E}_0\!\!\!\rhoo(-E^\prime)dE^\prime\right)\ , \quad E<0\ ,\nonumber\\
&=& - \frac{1}{8\pi E}\exp\left(-V_{\rm eff}(E)\right)\ ,
\eea
where, as will be discussed below, $V_{\rm eff}(E)$ is the leading (saddle point) effective potential for one eigenvalue in the $E<0$ regime~\cite{David:1990ge,David:1991sk}. These expressions, derived elegantly in a different manner in ref.~\cite{Saad:2019lba}, are an informative guide to what to expect non-perturbatively,  and they work well  for $E\gg\hbar$, but it must be stressed that they are only the first rough draft of the non-perturbative physics. They break down near the origin, are lacking various instanton contributions,  of course are based only on the leading piece $u_0(x)$, and---crucially---are in desperate need of knowledge of the trans-Fermi regime. All these aspects can be captured in the full framework under discussion.
 
\subsection{Non-Perturbative Data II: String Equations} 
\label{sec:string-equations}

Here is the heart of the matter this paper addresses. The double-scaled matrix model's governing equation for $u(x)$ is defined on the {\it whole real line} $x\in\mathbb{R}$.  This should not come as a surprise:  The index $n$ on the orthogonal polynomials $P_n(\lambda)$ does not stop at $N$, hence $X=n/N$ does not stop at 1. Correspondingly   $x$ does not stop at the Fermi level $\mu$. So there are data about the model coming from the $x\,{>}\,\mu$ or ``trans-Fermi'' region that  helps constrain what the complete physics is. So while perturbation theory to all orders in $\hbar$ can be described as expansion in the   $x\leq\mu$ regime,  it should be evident now that the matrix model is  incompletely described without a specification of the behaviour in the trans-Fermi regime. This is the key to understanding the consistent non-perturbative physics that completes the model.  To put it more boldly: 

\begin{itemize}[leftmargin=*]
\item{\it Any non-perturbative definition of the model should be equivalent to a specification of $u(x)$'s behaviour that fully extends into the trans-Fermi region.}
\end{itemize}

\noindent To turn it around, a key argument of this paper is that any proposed non-perturbative completion of JT gravity is likely problematic if it is inequivalent to such a specification. That would mean that it is either inconsistent, or perhaps has gone beyond what a matrix model can describe.\footnote{There is in principle nothing wrong with going beyond a matrix model to find  a non-perturbative completion, but this then comes with the question as to what the guiding physical principle is for accepting one completion over another.}

The  defining equation that comes from the double scaling limit of the (basic) Hermitian matrix model is 
\be
\label{eq:simple-string-equation}
{\cal R}=0\ ,\quad {\rm where}\quad {\cal R}\equiv \sum_{k=1}^\infty t_k R_k[u] +x\ ,
\ee
where the $R_k$ are the Gel'fand-Dikii polynomials~\cite{Gelfand:1975rn} in~$u(x)$ and its derivatives, normalized so that $R_k{=}u^k{+}\cdots$.  Their terms contain only even numbers of $x$-derivatives, with a power of $\hbar$ for each $\partial/\partial x$. Perturbation theory comes from first taking $\hbar{\to}0$ whence the equation for $u_0(x)$, the leading part of $u(x)$, is equation~(\ref{eq:leading-R0}). Expanding about this to include higher order~$\hbar$ corrections is then equivalent to an $\hbar/x$ expansion for large negative~$x$, confirming that genus perturbation theory needs only knowledge of the $x<\mu$ behaviour.

\begin{figure}[t]
\centering
\includegraphics[width=0.45\textwidth]{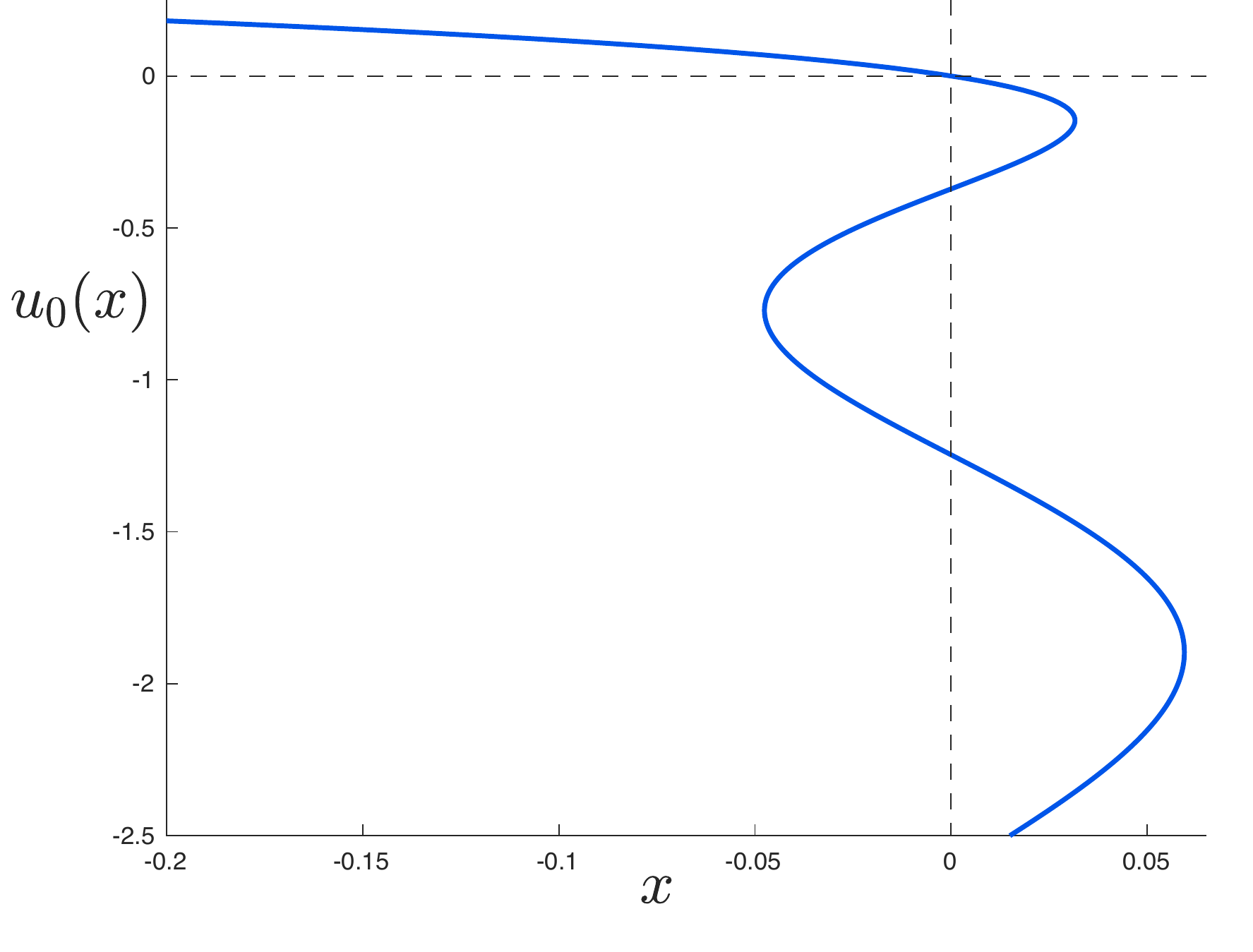}
\caption{\label{fig:multivalued-u} The solution for $u_0$ of equation~(\ref{eq:leading-u}), showing its multi-valuedness.}
\end{figure}

There are two ways of seeing the non-perturbative problems of the model. The most straightforward sign is that, some way above $x{=}\mu$, (recall $\mu{=}0$) the leading perturbative solution~(\ref{eq:leading-u}) for $u_0(x)$ reverses its direction   and folds back on itself in  {\it multivalued} behaviour. See figure~\ref{fig:multivalued-u}.  

%
As pointed out in ref.~\cite{Johnson:2020lns}, this does not make sense for a potential of  a quantum mechanics, but more damningly,  there is simply no way that a fully non-perturbative solution of a differential equation could unambiguously reduce to a multi-valued $u_0(x)$ in the limit of $\hbar{\to}0$. The result is that equation~(\ref{eq:simple-string-equation}) does not have a solution for $u(x)$ that can serve as a non-perturbative completion.  
It might seem odd that this could happen, but in fact it is quite natural.

\begin{figure}[t]
\centering
\includegraphics[width=0.48\textwidth]{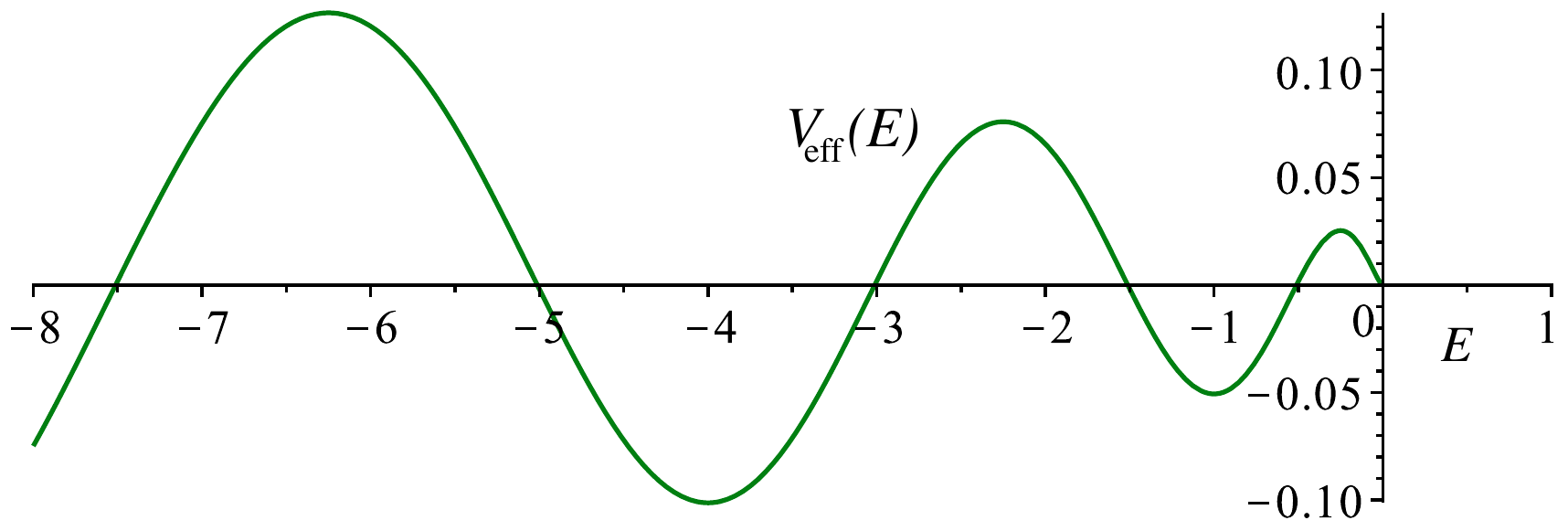}
\caption{\label{fig:effective-potential} The effective potential $V_{\rm eff}(E)$ for one eigenvalue from a semi-classical analysis of JT gravity.}
\end{figure}

To see why, it is useful to turn to the other, related sign of a problem. This comes from the following semi-classical analysis~\cite{David:1990ge,David:1991sk,Shenker:1990uf,Saad:2019lba}: A standard semi-classical  (leading large~$N$) result already seen in equation~(\ref{eq:semi-classical-np-2}) is that ($2\pi$ times) the disc spectral density, $\rhoo(E)$, continued to negative $E$, is the derivative of the effective potential $V_{\rm eff}(E)$ seen by one eigenvalue:
\bea
\label{eq:effective-potential}
&&\frac{d V_{\rm eff}(E)}{dE} = \frac{\sin(2\pi\sqrt{-E}}{2\pi\hbar}\ ,\,\,\, {\it i.e.,}\\
 && V_{\rm eff}(E) = \frac{1}{4\pi^3\hbar}\left(\sin(2\pi\sqrt{-E}) {-} 2\pi\sqrt{-E}\cos(2\pi\sqrt{-E})\right)\ , \nonumber
\eea
which is plotted in figure~\ref{fig:effective-potential}. So an eigenvalue sees an infinite number of additional minima to which it can tunnel to produce a new configuration. This  invalidates the saddle point solution that perturbation theory is being developed about. The leading instanton associated with this is  $\exp(-\frac{1}{4\pi^2\hbar})$, which comes from evaluating $V_{\rm eff}$ at the top of the first barrier, at $E{=}{-}\frac14$. The understanding of the role of the quantum mechanics described above makes  the  direct connection between the wiggles in $V_{\rm eff}(E)$ and the wiggles in $u_0(x)$.\footnote{The analytic continuation of~(\ref{eq:integrate-momentum}) to $E{<}0$, gives multiple answers for the integral since $u_0(x)$ develops multiple branches as $x$ approaches $\mu(=0)$, translating into the multiple branches of $V_{\rm eff} $ for a given $E$.}

The function $u(x)$ defined the family of orthogonal polynomials that builds the configuration of interest for studying a gravity dual. The fact that it becomes  mulitivalued reflects the ambiguity in how to extend to the trans-Fermi regime in order to determine where the eigenvalues live in the $E{<}0$ energy regime. (Note that an important part of the dictionary is necessarily that $E{<}0$ states are the domain of the trans-Fermi regime and hence pertain to non-perturbative physics. This is because $u_0(x)$  is positive in the Fermi sea regime and so  $E{<}0$ is classically forbidden and can only contribute to the Fermi sea as tunneling effects {\it i.e.,} non-perturbative in the $\hbar$ expansion.)   The string equation was derived by implicitly assuming only that the solution has all the energies in the one connected Dyson droplet (often called a ``one-cut'' solution) and so it is no surprise that it fails to extend $u(x)$ to the trans-Fermi regime.

While there are likely new vacua of the model to be found\cite{Dalley:1991zr}, representing multi-cut solutions,  it should not be forgotten that the primary goal was to seek a description of  JT gravity, and this will remain the focus instead of chasing after the (interesting) new solutions. This means a stable system that has leading  $\rhoo(E)$ as in equation~(\ref{eq:leading-density}), a single-cut solution, should be found. 
The conclusion must be that string equation~(\ref{eq:simple-string-equation}) can be  correct only perturbatively, but for non-perturbative physics something else must be sought.

A natural way forward is to realize that the Hermitian matrix model equation ${\cal R}{=}0$  can be  embedded into a more general equation:
\be
\label{eq:big-string-equation}
(u-\sigma){\cal R}^2 - \frac{\hbar^2}{2}{\cal R}{\cal R}^{''}+\frac{\hbar^2}{4}({\cal R}^{'})^2 = \hbar^2\Gamma^2\ ,
\ee
where for now, $\Gamma$ will be set to zero.
This equation's origins (for $\Gamma{=}0$) can be thought of in at least two ways. With $\sigma{=}0$, it  was first found~\cite{Morris:1991cq,Dalley:1992qg}, by studying  complex matrices $M$ with an action that depends only on $MM^\dagger$, so it is better here to think of them as defining a system of random positive Hermitian matrices. (See footnote~\ref{foot:remark} on page~\pageref{foot:remark} for more on this.) The case of non-zero $\sigma$ is a  model  of random Hermitian matrices with $\sigma$  as their lowest energy, as can be shown~\cite{Dalley:1991xx}  by {\it e.g,} performing the double-scaling limit with a ``wall'' at scaled position $\sigma$.\footnote{The complex matrix model interpretation is much more natural in the context of a certain kind JT supergravity model that belongs in the $(2\Gamma+1,2)$ Altland-Zirnbauer classification discussed in ref.~\cite{Stanford:2019vob}. Then the complex matrix $M$  correlates nicely with the Type 0A  realization of the 2D supersymmetry, and an unambiguouus non-perturbartive completion results for each case~\cite{Johnson:2020heh,Johnson:2020exp}.}  In fact, naively sending $\sigma\to-\infty$ in the equation picks out the Hermitian matrix model string equation ${\cal R}{=}0$ again. Intuitively, this is the right equation to use since it has three attractive characteristics:
\medskip

\begin{itemize}[label={\raisebox{2pt}{\tiny\textbullet}},leftmargin=*]
\item {${\cal R}{=}0$ is a perturbative solution in the $x\leq\mu$ regime, therefore reproducing JT gravity perturbation theory to all orders.}

\item { It does not allow the spectrum to run off to arbitrarily negative values. For finite  $\sigma$  the equation defines a non-perturbatively self-consistent system that does not allow for tunneling effects to arbitrary negative $E$ that invalidate the assumptions of perturbation theory.}
\item { It is derived from a random Hermitian matrix model, which is what was used to define perturbation theory. This is  a nice bonus.}
\end{itemize}

Following the above logic of why the simpler string equation failed, it is clear why this has every chance to succeed self-consistently. Just one small extra assumption was made in addition to assuming a single-cut solution (required for JT gravity) and that is that there is some lowest energy for which such a single-cut situation is consistent.
As will become clear shortly, simple criteria will show that to obtain JT gravity, the allowed range for $\sigma$ is not chosen by hand, but self-consistently by the  equation (and hence the Hermitian matrix model) itself. For example it will become clear that the suggested minimum value  $E{=}{-}\frac14$ from the semi-classical analysis is corrected upwards significantly. (This is not too surprising, as it is a semiclassical analysis after all.)

\begin{figure}[t]
\centering
\includegraphics[width=0.45\textwidth]{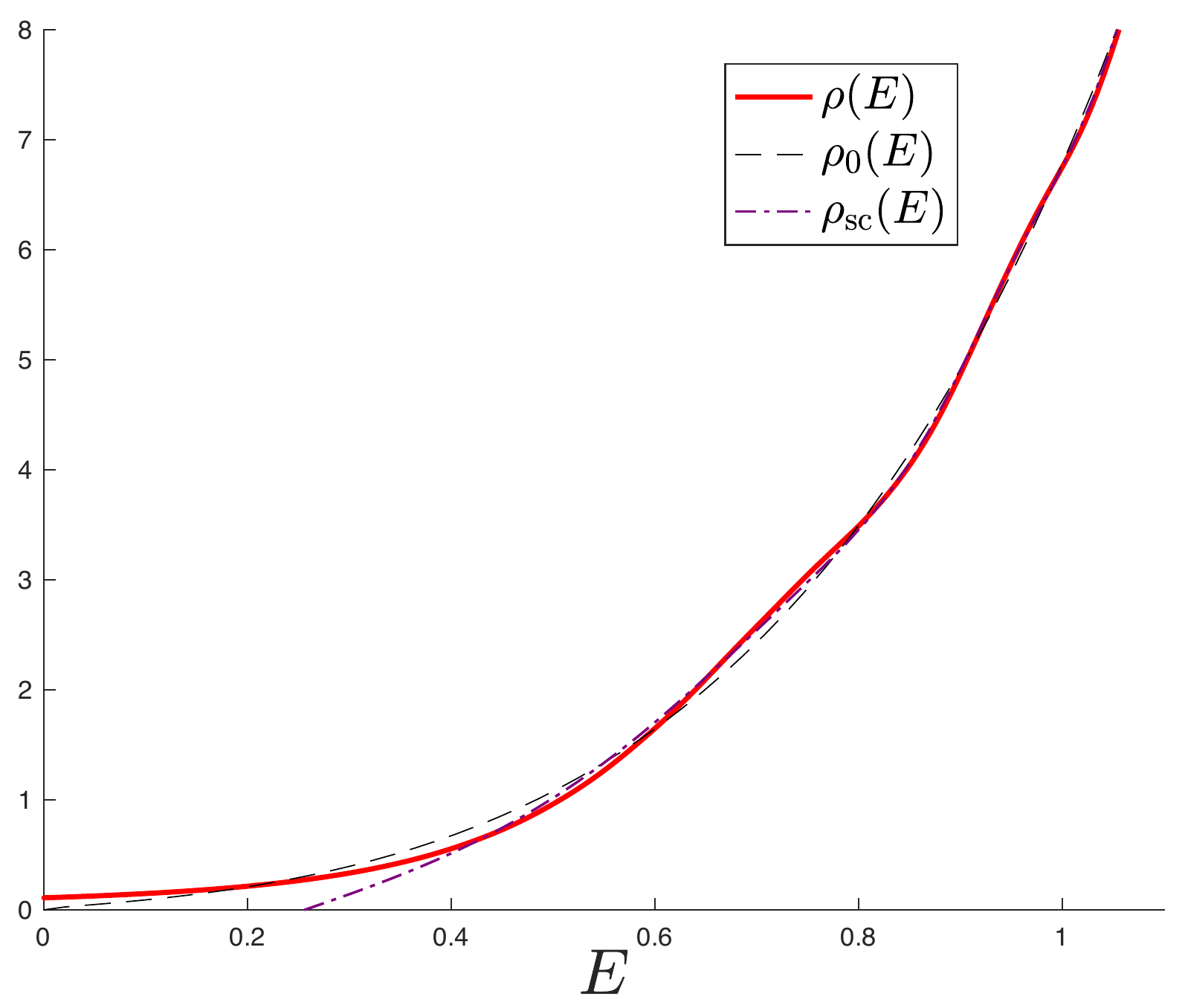}
\caption{\label{fig:full-density-vs-semi-classical} The full non-perturbative spectral density $\rho(E)$ (solid red), the leading disc result $\rho_0(E)$ and the semi-classical estimate $\rho_{\rm sc}(E)$. Here $\hbar{=}1$.}
\end{figure}

Ref.~\cite{Johnson:2019eik}'s non-perturbative completion  used string 
equation~(\ref{eq:big-string-equation}) with $\sigma=0$,  as follows: At $\hbar=0$, a continuous, single-valued extension of $u_0(x)$ to the whole real line is simply: 
\bea
\label{eq:prototype-defintion}
{\cal R}[u_0(x)] &=& 0\quad x\leq0\ ,\nonumber\\
u_0(x)&=&0\quad x>0\ .
\eea 
Solving the full equation~(\ref{eq:big-string-equation}) with this as the boundary condition then gave a self-consistent, stable, non-perturbative defintion of JT gravity with the spectrum ending at $E{=}0$. The result is shown in figure~\ref{fig:full-density-vs-semi-classical}, alongside the leading disc spectral density, around which it shows marked oscillations. It is  strongly corrected by non-perturbative effects near $E{=}0$, ending up at some finite $\rho(0)$ at $E{=}0$. (Also shown in the figure is  ref.~\cite{Saad:2019lba}'s semiclassical estimate~(\ref{eq:semi-classical-np-1}) for non-perturbative effects, which does pretty well for large $E$ but as anticipated goes off the rails  toward small $E$.) 
This completion, built on equation~(\ref{eq:prototype-defintion}), was of course a fine  choice.  Much has been learned from it, and computed using it, as recalled in the Introduction. However, there are other choices that  quite readily arise, and all the tools and intuition discussed so far are primed to examine and compare them.

\section{A Family of Completions}
\label{sec:consistency-conditions}
\subsection{Consistency Conditions}
\label{sec:consistency-conditions-sub}
To understand the family of non-perturbative definitions to follow,  it is worth summarizing the criteria needed to obtain a completion of JT gravity, while also preserving perturbation theory:

\medskip
\noindent As $\hbar{\to}0$, any non-perturbative solution for $u(x)$ {\it must...}
\begin{itemize}[leftmargin=*]
\item {\it ...reduce to a $u_0(x)$ that is continuous and  single-valued on the real line. (Relaxing the continuity requirement will be discussed below.)}

\item {\it ...yield ${\cal R}_0=0$, {\it i.e.,} equation~(\ref{eq:leading-u}),  in the $x{<}0$ regime, since this gives the  Fermi sea data that enables  the matrix model to yield the correct perturbation theory.}
\end{itemize}

\begin{figure}[t]
\centering
\includegraphics[width=0.45\textwidth]{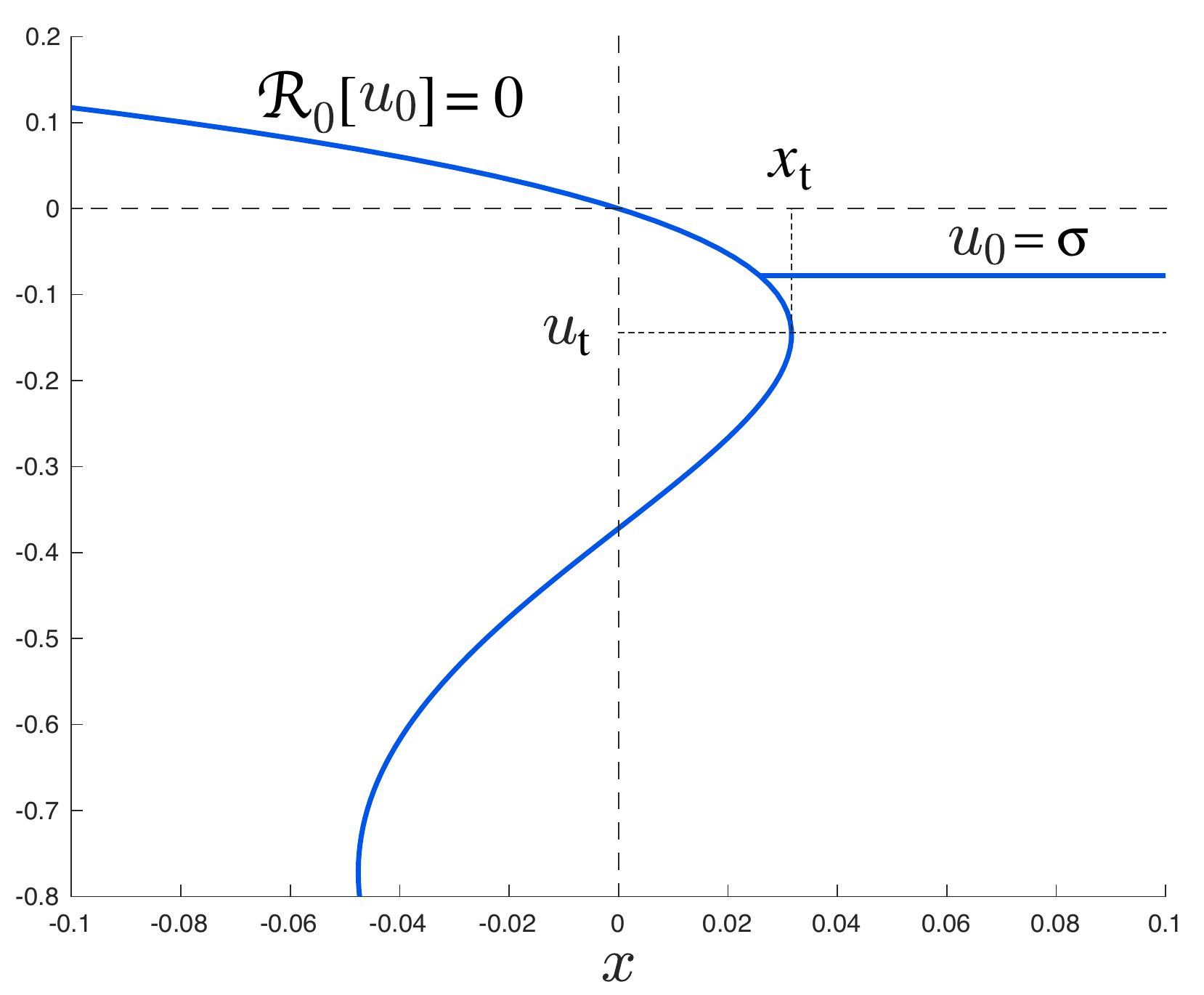}
\caption{\label{fig:sigma-scheme} A family of  single-valued continuous leading order solutions to equation~(\ref{eq:big-string-equation}) can be made by gluing the solutions ${\cal R}_0{=}0$ and $u_0=\sigma$ as shown. Consistency allows $|\sigma|$ to be as small as zero and as large as $u_{\rm t}$, defined by where the left curve begins to turn.}
\end{figure}

While~(\ref{eq:prototype-defintion}) is a consistent choice that fits with the above conditions, it is clear that there is a larger class of solutions available based on boundary condition:
\bea
\label{eq:broader-defintion}
{\cal R}_0 &=& 0\quad x\leq {\hat x}\ ,\nonumber\\
u_0(x)&=&\sigma\quad x\geq{\hat x}\ , \nonumber\\
\hskip-0.2cm{\rm with }\quad {\hat x}\in(0,x_{\rm t})&&{\rm and}\quad \sigma\in(0,u_{\rm t})\ ,
\eea 
where $x_{\rm t}$ is the location at which $u_0(x)$ first begins to turn around, at value $u_{\rm t}$. Their values are given below in equation~(\ref{eq:turning}). The position ${\hat x}$ can be anywhere between the Fermi surface value at $x{=}0$ and the turnaround $x_{\rm t}$, and $\sigma$ is the value  $u_0({\hat x})$.    A sketch is given in figure~\ref{fig:sigma-scheme}.

Once the above extension to the leading solution is used in the full string equation, the methods already demonstrated in ref.~\cite{Johnson:2020exp} for finding a good non-perturbative solution go through as before. The spectral problem  can readily be solved. The chosen value of $\sigma$ sets the lowest energy of the spectrum.  

It is very important to appreciate this this does {\it not} correspond to taking the double-scaled Hermitian matrix model spectrum and simply truncating it at $\sigma$, throwing away the lower energies. This restriction of the energies to $E\in(\sigma,+\infty)$, for some $\sigma{<}0$,  is a self-consistent model where, in effect, the spectrum responds to the presence of the bound. This is already clear from considering the quantum mechanics toolbox~(\ref{eq:schrodinger}).  The spectrum of wavefunctions obtained for $u(x,\sigma^\prime)$ is {\it not} a subset of those for $u(x,\sigma^{\prime\prime})$, where $\sigma^\prime>\sigma^{\prime\prime}$. This will be made even more explicit with examples in the next section (see the inset of figure~\ref{fig:many-spectra}).

It is interesting to look at the case where $\sigma{=}u_{\rm t}$, the spectrum with the lowest  energy possible that fits the above criteria (so far). 
Using equation~(\ref{eq:leading-u}) yields $dx/du_0=-\frac12 I_0(2\pi\sqrt{u_0})=-\frac12J_0(2\pi\sqrt{|u_0|})$, (because $u_0$ is negative in this region), and so the values of $u_0$ and $x$ at the turning point are:
\begin{eqnarray}
\label{eq:turning}
u_{\rm t} &=& -\left(\frac{j_{01}}{2\pi}\right)^2\simeq -0.1464898\ , \nonumber\\
 x_{\rm t} &=& -\frac{j_{01}}{(2\pi)^2}J_1(j_{01})\simeq - 0.0316238\ ,
\eea  where $j_{01}$ is the first zero of the Bessel function $J_0$. The value $u_{\rm t}$ is the lowest energy consistently allowed by continuity (if not adding D-branes---see Section~\ref{sec:D-branes}). It compares interestingly to the position of the  peak of the first tunneling barrier, from the semi-classical analysis of equation~(\ref{eq:effective-potential}), which is at $E=-\frac14$, {\it i.e.,} further to the left. In other words, this non-perturbative completion, although it now allows states with $E<0$,  still truncates the spectrum's tail well before the semi-classically predicted danger point.

New possibilities open if the continuity requirement of the consistency conditions is relaxed. Evidence for this is the fact that  numerical exploration of individual minimal models with decreasing $\sigma$ in  the early results of ref.~\cite{Johnson:1992wr} showed the existence of  solutions that  are not continuous in the  $\hbar{\to}0$ regime. In other words, the $u_0{=}\sigma$ component  is so low that it cannot connect to the upper branch of the ${\cal R}_0{=}0$ sector. For the simplest such models explored, the solutions (for $k$ even) developed increasingly deep wells as $\sigma$ is lowered.The wells might become deep enough to support bound states, which  may well be precursors of multi-cut solutions to which the instability of the simpler equation~(\ref{eq:simple-string-equation}) points. Direct exploration of  values of $\sigma$ below $u_{\rm t}$ in this JT case seems to find solutions, but they have proven difficult to fully explore, being at the edge of what seems currently possible  with the techniques being employed.  The core point is that such solutions extend the range of $\sigma$ for which there are viable non-perturbative completions. Further work on these matters would be interesting.

\subsection{Explicit Examples}
\label{sec:explicit}
A beauty of this formalism is how readily  explicit results can be (with care) computed and displayed. The methods for solving equation~(\ref{eq:big-string-equation}) are thoroughly discussed in ref.~\cite{Johnson:2020exp}, and will not be repeated here. (The value $\hbar{=}1$ will be used throughout for illustration. It is also the value at which non-perturbative effects are maximized.) Figure~\ref{fig:many-potentials} shows a few examples (setting $\hbar{=}1$) of the  solution for $u(x)$ (using a $k{=}7$ truncation, in the scheme of ref.~\cite{Johnson:2020exp}) obtained for five values of $\sigma$, for illustration purposes, with the first being the already known case $\sigma{=}0$, and the last being at the other extreme of the range discussed above, $\sigma{=}u_{\rm t}{=}{-}(j_{01}/2\pi)^2$. 
\begin{figure}[t]
\centering
\includegraphics[width=0.5\textwidth]{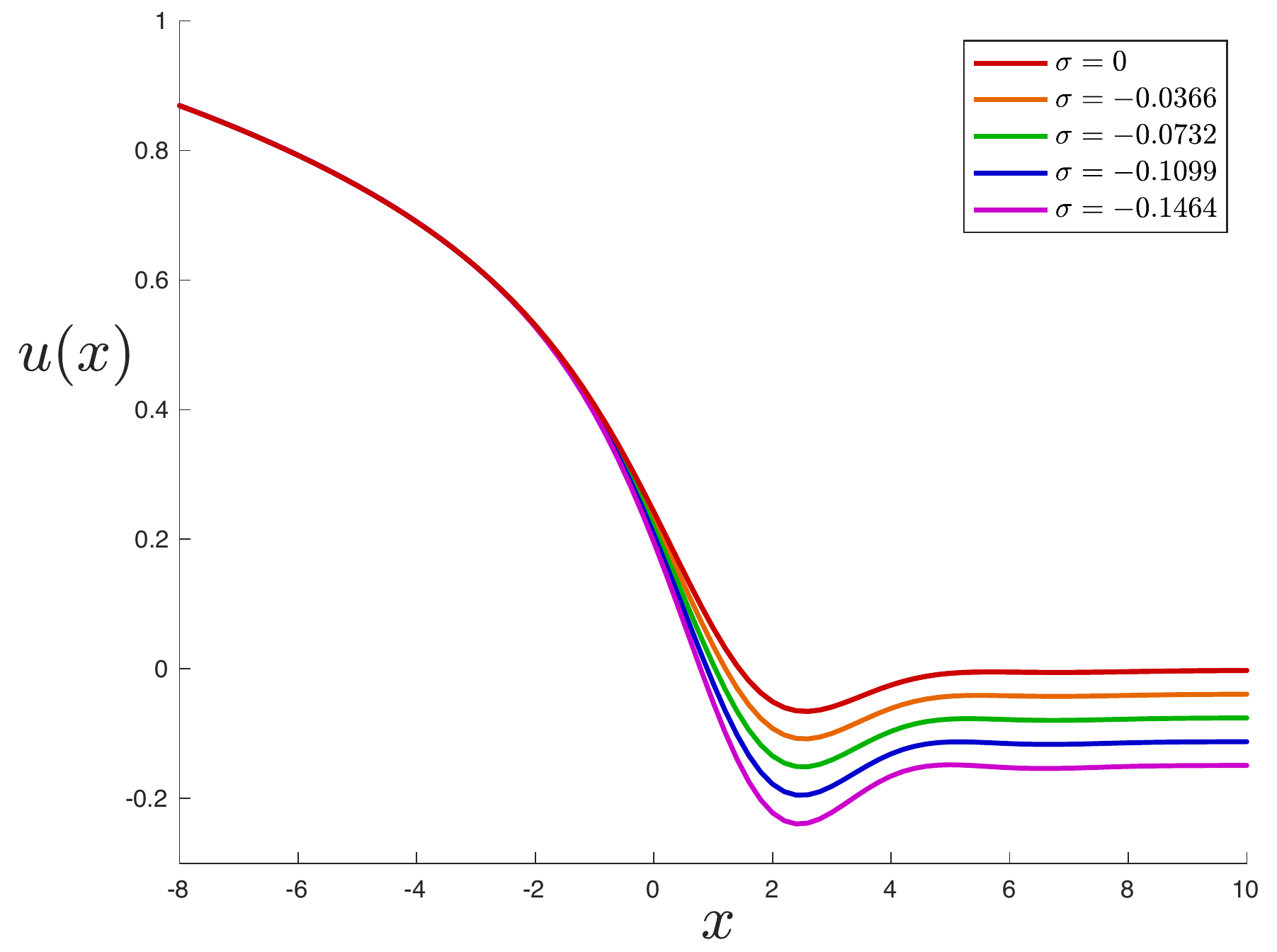}
\caption{\label{fig:many-potentials} The  potentials for various values of $\sigma$, from 0 (highest) down to and $\sigma{=}u_{\rm t}\simeq-0.1464$ (lowest). Here $\hbar{=}1$.}
\end{figure}

As emphasized earlier, being able to quantitatively compare different completions of JT gravity is an advantage of having them in a single framework, and this can be done now. 
Consulting figure~\ref{fig:many-potentials}, the non-perturbative parameter $\sigma$'s nature is already intuitively clear
given the dictionary outlined earlier: It has no effect
deep in the Fermi region, the domain of perturbative data,
only beginning to have an effect nearer the Fermi level (the smaller~$|x|$ Fermi region differences are somewhat amplified with  $\hbar{=}1$), and
 is fully manifest in the trans-Fermi region as 
a shift in~$u(x)$ away from zero by $\sigma$ as $x{\to}\infty$. Such a shift translates, in a complete solution, to something exponentially small in the far Fermi region where physical quantities are computed perturbatively (recall, as an expansion in small $\hbar/|x|$). The effects of such a shift can be readily estimated when $\hbar$ is small, where it is then more akin to an overall energy shift by $\sigma$ in the leading perturbative potential, with a resulting instanton form for the action: $\e^{-V_{\rm eff}(\sigma)}$ where $V_{\rm eff}(E)$ is given in equation~(\ref{eq:effective-potential}). This, then, is the leading estimate of how the different completions compare to each other.

As a contrast, this can be compared to a perturbative shift of $u(x)$ by some amount $\tau$, which is to say, it shows up deep in the  Fermi sea region as a shift of that amount.  The expectation value of a loop of length~$\ell$, denoted $\langle W(\ell)\rangle{=} \int^\mu \langle x| \exp\{ -\ell(-\hbar^2\partial^2_x+u(x))\}|x\rangle dx$, (the JT partition function~(\ref{eq:JTpartfun}) is one, with $\ell{=}\beta$), simply gets multiplied
by a factor of $\e^{-\ell\tau}$. This weighting loops of length $\ell$  would indicate the action of a perturbative boundary operator. Instead, $\sigma$ will be seen to add a non-perturbative contribution to the boundary length operator, as will be discussed in Section~\ref{sec:dynamical-boundaries}.

The $u(x)$ of each completion can  be used to compute a spectral density, using the methods of ref.~\cite{Johnson:2020exp}, and compared. As might be expected, they are extremely similar, especially at higher $E$, with slight differences setting in at lower energies. See figure~\ref{fig:many-spectra}). In particular,  around $E=0$ (see inset) the density's value is lower for  smaller~$|\sigma|$.

%

%
\begin{figure}[t]
\centering
\includegraphics[width=0.48\textwidth]{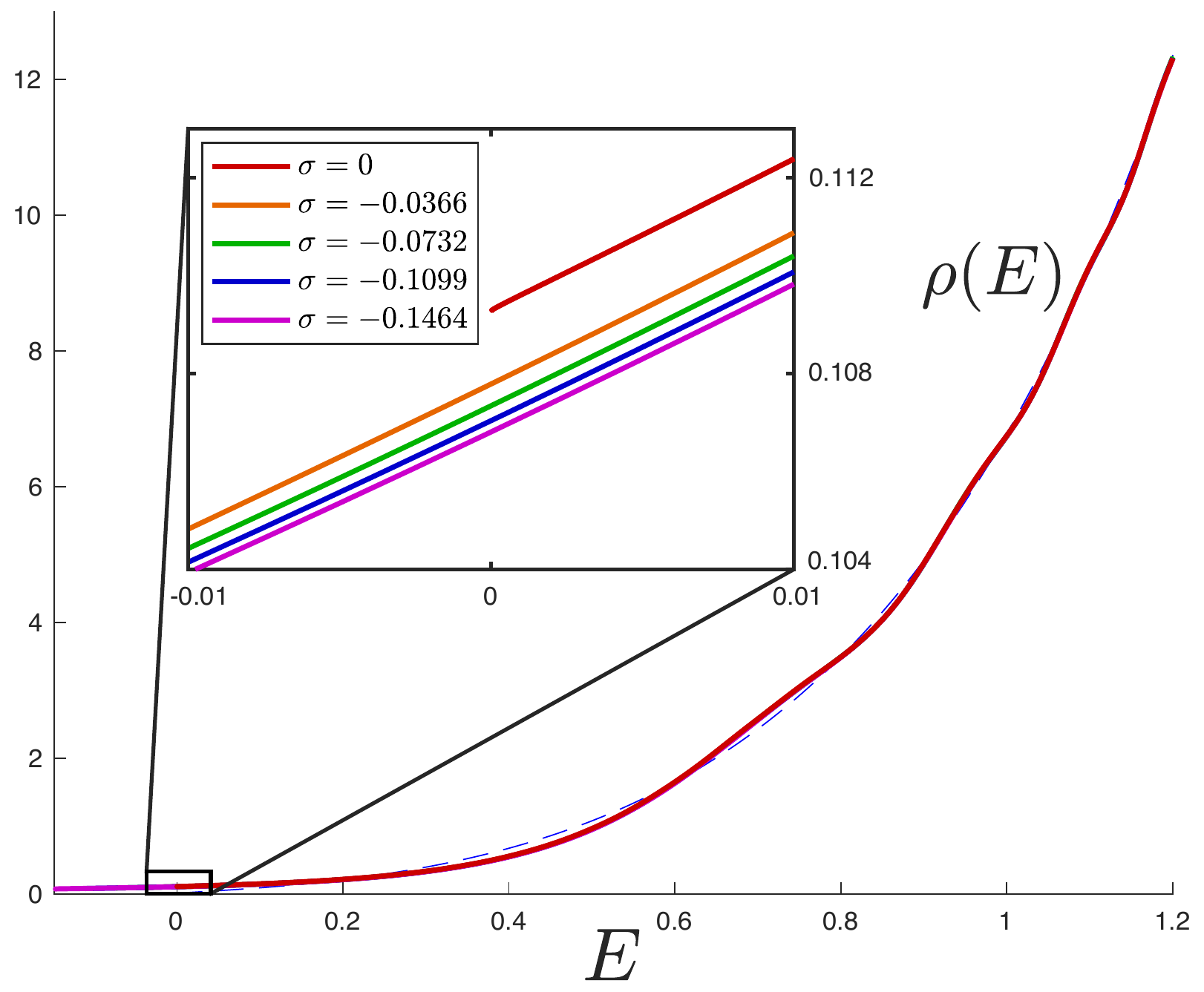}
\caption{\label{fig:many-spectra} The  non-perturbative JT gravity spectral densities for  various values of $\sigma$, from 0 (highest) down to and $\sigma{=}u_{\rm t}\simeq-0.1464$ (lowest). The inset confirms that they have  significant differing behaviour near $E=0$, which must be the case for consistency (see text). Here $\hbar{=}1$.}
\end{figure}
%

A key characteristic quantity to compute for a non-perturbative completion is the  two-point correlator for a pair of energy levels $E$ and $E^\prime$, normalized to unity and denoted $R_2(E,E^\prime)$ so that it gives the probability of finding an energy level at $E^\prime$ given an energy level is at~$E$. The structure of this quantity is a key feature of a model, encoding  non-perturbative physics that controls other quantities, such as the spectral form factor. This is readily computed, and it is interesting to see how if the different  sample different non-perturbative completions would show many differences.  For the case of $\hbar=1$ (where the effects would be most marked) the differences were of order smaller than $10^{-4}$. Figure~\ref{fig:two-point} shows this as best as it is able (placing one energy at the origin), with all five curves coincident to well within the thickness of the plotting line. An extreme magnification to examine four points on each curve shows the different curves slightly separated out.   

\begin{figure}[b]
\centering
\includegraphics[width=0.5\textwidth]{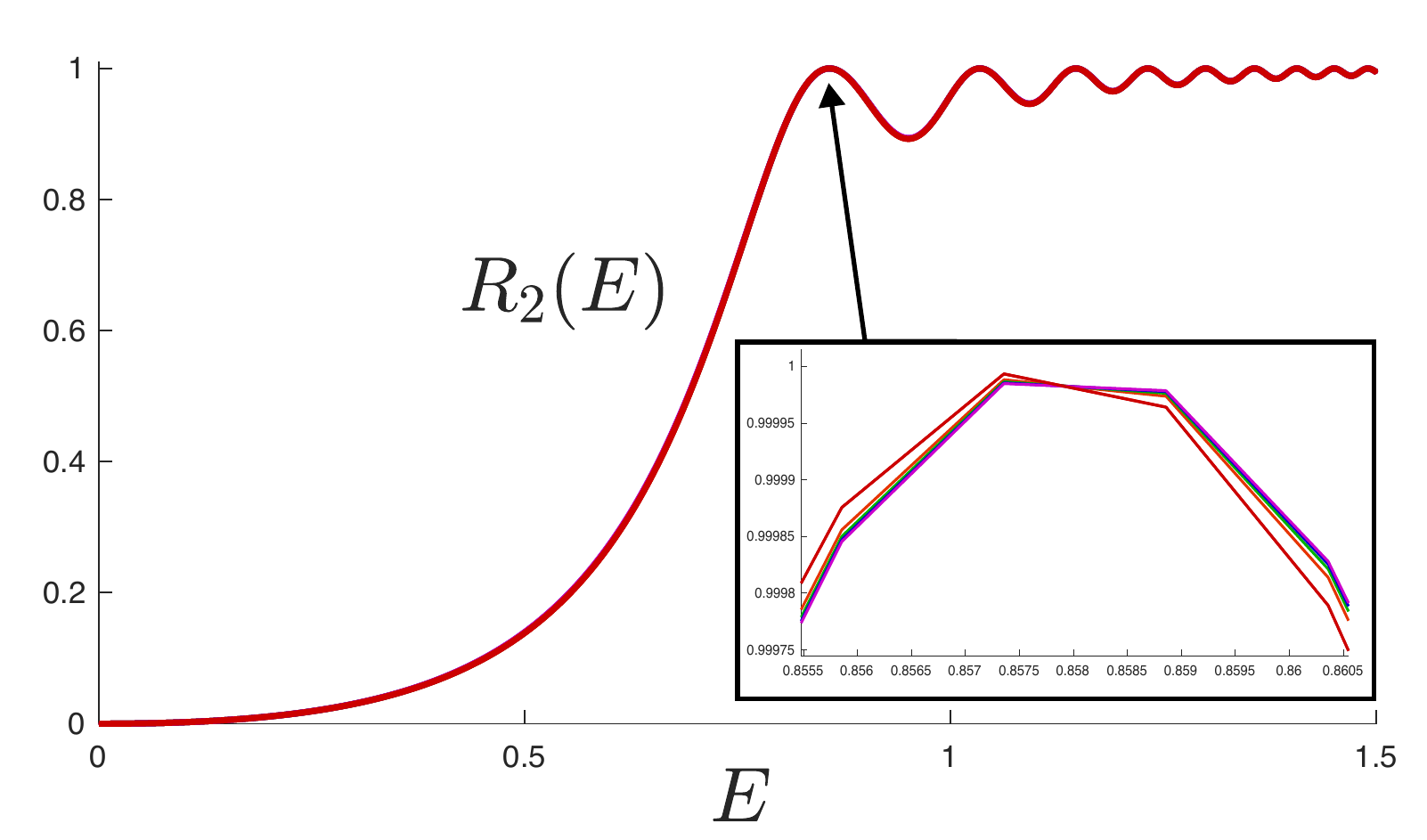}
\caption{\label{fig:two-point} The  JT gravity two-point function for five different non-perturbative completions.  The inset shows how little the physics changes between completions (see text). The colour coding matches the previous figure.  Here $\hbar{=}1$.}
\end{figure}

With the full wavefunctions $\psi(E,x)$ in hand, numerous other quantities are readily computable, including (following ref.~\cite{Johnson:2021zuo}'s use of the Fredholm determinant) the individual microstate statistical distributions.  For the special case $\sigma{=}u_{\rm t}$, the microstate results are shown in figure~\ref{fig:JT-microstates}. In a non-standard notation (compared to  the random matrix literature~\cite{Meh2004,ForresterBook}), $F(n,s)$ is the probability density function for energy $s$ for the $n$th level while $E(n,s)$ are cumulative density functions (dashed lines). 

\begin{figure}[t]
\centering
\includegraphics[width=0.48\textwidth]{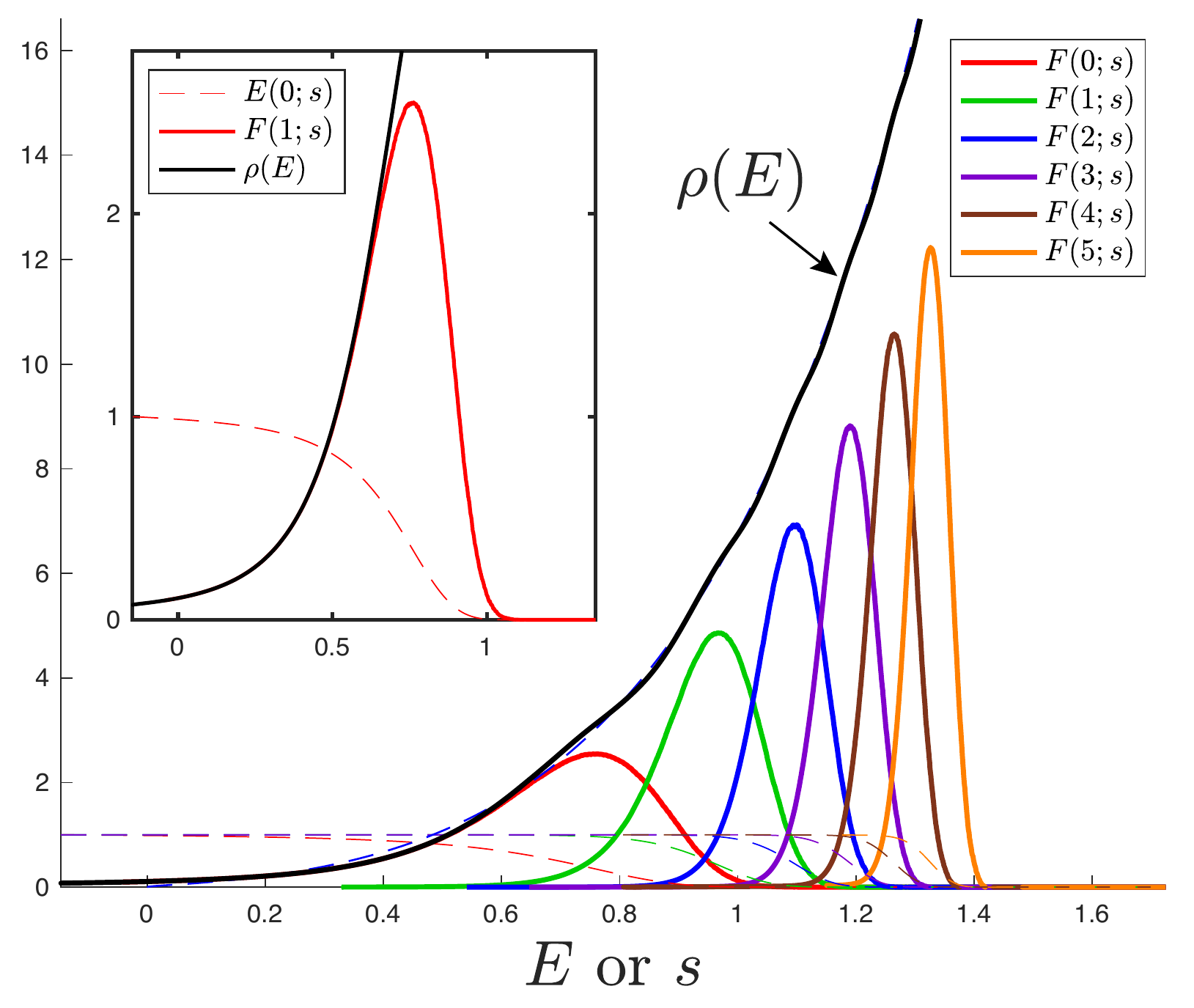}
\caption{\label{fig:JT-microstates} The  JT gravity ``microstate'' spectrum for the case $\sigma{=}u_{\rm t}{\simeq}-0.1464$, showing the probability distributions of the fist six energy levels.  Here $\hbar{=}1$.}
\end{figure}

 As with the $\sigma{=}0$ case, the full quenched free energy $F_Q(T){=}{-}T\langle\log Z(T)\rangle$ can be computed in each case too. The results look similar to the prototype case computed and displayed in ref.~\cite{Johnson:2021zuo}  with a slightly smaller value for the $F_Q(0)$, consistent with the fact that the average ground state energy shifts  slightly down with decreasing~$\sigma$. (It is only a slight shift since the population of energies in the $E<0$ regime is exponentially small.)

\bigskip


\section{Background, or ``End-of-the-World'', Branes}
\label{sec:D-branes}
There has been recent discussion in the literature about inserting ``End-of-the-World'' branes into matrix models of JT gravity. (Note that they are part of the previously mentioned non-perturbative proposal of ref.~\cite{Gao:2021uro}.)  These are nothing but D-branes, and the technology for introducing them as background objects into the current matrix model setting has been well understood for some time now. They are straightforward to incorporate  into the consistency conditions  of Section~\ref{sec:consistency-conditions}, and it is illuminating to see the results.

Again,  in this framework everything can (and should)  be discussed in terms of their effects on $u(x)$. Step one is as follows: Some number, $\Gamma$ of the branes associated with parameter $\sigma$ can be introduced into the model by turning on $\Gamma$ in equation~(\ref{eq:big-string-equation}). In fact, there is no step two. That's really all there is to it, and such background branes have been studied in just this way in the literature in {\it e.g.} refs.~\cite{Dalley:1992br,Johnson:1994vk,Johnson:2004ut}.
Here, $\sigma$, which is a position in the Dyson gas, or an energy in the auxiliary quantum mechanics, is the boundary cosmological constant of the background D-branes introduced.

 The point is that the string equation~(\ref{eq:big-string-equation}) naturally has the potential for describing background D-branes  built in, as extensively described in ref.~\cite{Johnson:2004ut}, where many effects of~$\Gamma$ on the closed string physics was worked out.  This ``naturalness'' follows from a number of features. One of the most direct was derived in refs.~\cite{Itoh:1992np,Johnson:1994vk}, where it was shown how the equation and the D-brane physics it describes follows from performing a redefinition of the closed string couplings. This will be discussed more in Section~\ref{sec:dynamical-boundaries}.  Another feature is that in the Fermi regime the equation is perturbatively  equivalent to an early open string  model of Kostov~\cite{Kostov:1990nf}, where the Hermitian matrix model string equation is modified to ${\cal R} =2 \hbar\Gamma{\hat R}(u,\sigma)$, with ${\hat R}(u,\sigma)\equiv\langle x|({\cal H}-\sigma)^{-1}|x\rangle$ being the diagonal of the resolvent of ${\cal H}$. (It can be interpreted as inserting as a  background the determinant operator ${\rm det}(M-{\sigma})$ into the random matrix model for Hermitian $M$.\footnote{See also {\it e.g.}, refs.~\cite{Fateev:2000ik,Zamolodchikov:2001ah,Teschner:2000md,Martinec:2003ka,McGreevy:2003kb,Maldacena:2004sn} for more on D-branes in minimal string theories.}) Asking that the resolvent solves Gel'fand and Dikii's~\cite{Gelfand:1975rn} equation for it: $4(u{-}\sigma){\hat R}^2-2\hbar^2{\hat R}{\hat R}^{\prime\prime}+\hbar^2({\hat R}^{\prime})^2{=}1$, results in the string equation. Another way of seeing $\Gamma$ appear in this way (for $\sigma{=}0$) is that it can be derived from rectangular matrix models~\cite{Anderson:1991ku,Myers:1991akt,Lafrance:1993wy,Klebanov:2003wg}, where a system of $N{\times}(N+\Gamma)$ complex matrices $M$ is used, again with potential based on $MM^\dagger$. Then $\Gamma$ is directly interpretable as the number of ``flavours'' of quark inserted into the model.

Recent  studies of the effect of $\Gamma$ (using direct studies of  matrix ensembles~\cite{Johnson:2021rsh}, where $\Gamma$ is modeled using rectangular complex matrices) have confirmed that the presence of non-zero positive $\Gamma$ produces a repulsive effect on the other energy eigenvalues, pushing the spectrum to the right, while there is a degenerate exact ground state with multiplicity $\Gamma$. Here, things will be similar,   but with  the background branes  placed at $\sigma$, which can be non-zero. It is clear that the tree level equation is deformed by $\Gamma$ at the next order in perturbation theory, with $u_0$ solving: ${\cal R}_0 = 2\hbar\Gamma{\hat R}(u_0,\sigma)$.  

This produces a deformation of the defining equation for $u_0(x)$ away from the form discussed in the previous section. The result will be the movement of the location of the turning points  at which the dangerous multivaluedness begins to occur.  Positive $\Gamma$ pushes them to more negative energies, as will become clear in a moment. For generic~$\Gamma$ and small $\hbar$ it is a small correction, but when the right hand side is no longer small, this is a significant deformation. It is worth dialing up the effects of the term in order to understand the physics.

An old and familiar story (and a useful one) is to take a limit on the open string sector (D-branes) where their effect can be re-interpreted as a closed string background. This can be done here, with $\Gamma\to\infty$ while $\hbar\to0$ so that~${\tilde \Gamma}\,{=}\,\hbar\Gamma$ is held fixed.\footnote{This was first done on this equation (with a different motivation)  in ref.~\cite{Klebanov:2003wg}, as a means of arguing for the existence of smooth solutions for  $\Gamma{=}0$ by deformation from large $\Gamma$.} Then to leading order the  string equation once again becomes a matter of  algebra:
\bea
&&(u_0-\sigma){\mathcal R}_0^2={\tilde\Gamma}^2\ , \quad {\it i.e.,}\nonumber \\
&&(u_0-\sigma)\left({\frac{\sqrt{u_0}}{2\pi}I_1(2\pi\sqrt{u_0})}+x\right)^2={\tilde\Gamma}^2\ ,
\eea
which can  be read as a simple deformation by ${\tilde \Gamma}$ that smoothly connects the two branches~(\ref{eq:broader-defintion}) of  the leading solution. Alternatively,
\be
{\frac{\sqrt{u_0}}{2\pi}I_1(2\pi\sqrt{u_0})}+x=\pm\frac{\tilde\Gamma}{\sqrt{u_0-\sigma}}\ ,
\ee
where the sign of ${\tilde \Gamma}$ is made more manifest. It makes a difference in the theory, as will be clear in a moment.  It is useful to compute the turning points in $x$ again:
\be
-\frac{\partial x}{\partial u_0} = \frac12 I_0(2\pi\sqrt{u_0})\pm\frac12 \frac{\tilde\Gamma}{(u_0-\sigma)^{\frac32}}\ .
\ee
\begin{figure}[t]
\centering
\includegraphics[width=0.45\textwidth]{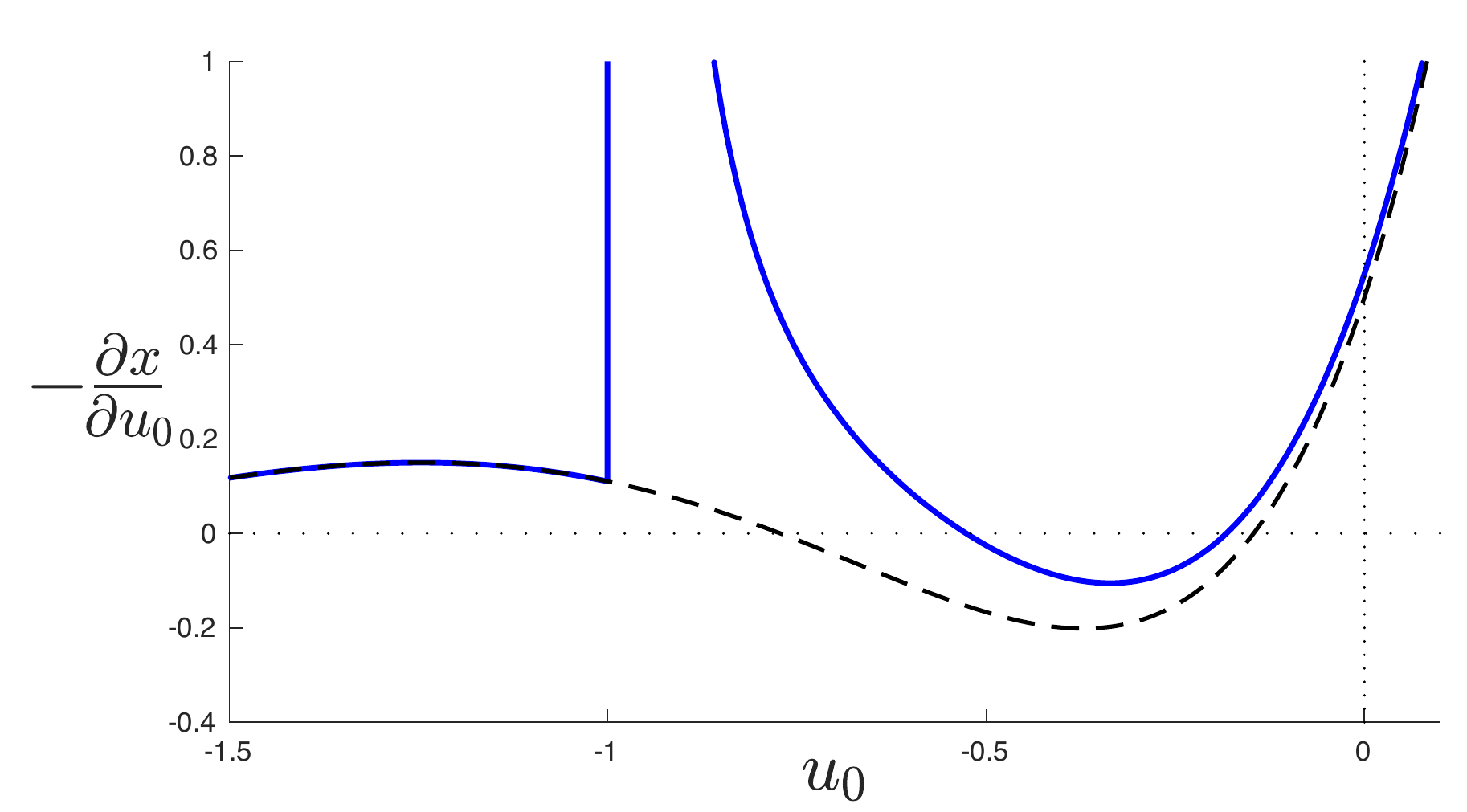}
\caption{\label{fig:JT-zeros} A plot of $\partial x/\partial u_0$ showing the deformation of the zeros marking the turning points of the leading string equation for $u_0(x)$ when ${\tilde \Gamma}{=}0.05$. The dashed line shows the case of ${\tilde\Gamma}{=}0$. Here, $\sigma{=}-1$. }
\end{figure}
A dominant feature is now the singularity at $u_0=\sigma$, which is simply where the $u_0(x)$ curve, starting in from negative $x$, eventually transitions to  the new branch that stretches out to infinity. But it can do that in two ways, depending upon the sign of $\tilde\Gamma$. A positive sign sends the curve to $x=+\infty$, and in fact $\tilde\Gamma$ has the effect of lifting the zeros that previously signalled the turning points. This can allow for $\sigma$s that are more negative than the limiting value seen with no background branes, depending upon the value of $\tilde\Gamma$.  This has an interpretation as the branes naturally being repulsive of the other energies in the spectrum, pushing them to more positive $E$ and hence making a larger portion  of the $E<0$ regime ``safe''.  As an example, ${\partial x}/{\partial u_0}$ for the case of $\tilde\Gamma{=}0.05$ and $\sigma{=}{-}1$ shown in figure~\ref{fig:JT-zeros}.  The dashed curve is the unperturbed ${\tilde \Gamma}{=}0$ case. Passing through zero indicates a turning point where multivaluedness sets in. In this example, even though the position of a pair of zeros has been deformed, it is not enough for the chosen value  $\sigma{=}{-}1$ to be a consistent choice. The multivaluedness of this $u_0(x)$ (it turns back, then again, before going off to $x{=}+\infty$) means that no non-perturbative completion can reduce to it as $\hbar\to0$. 
Turning up the value of~$\tilde\Gamma$ to 0.1 changes the situation. See  figure~\ref{fig:JT-zeros2}. There, the pair of zeros has merged and disappeared, meaning that the $u_0(x)$ is a consistent seed for a non-perturbative completion, smoothly going from left to right uneventfully.
\begin{figure}[t]
\centering
\includegraphics[width=0.45\textwidth]{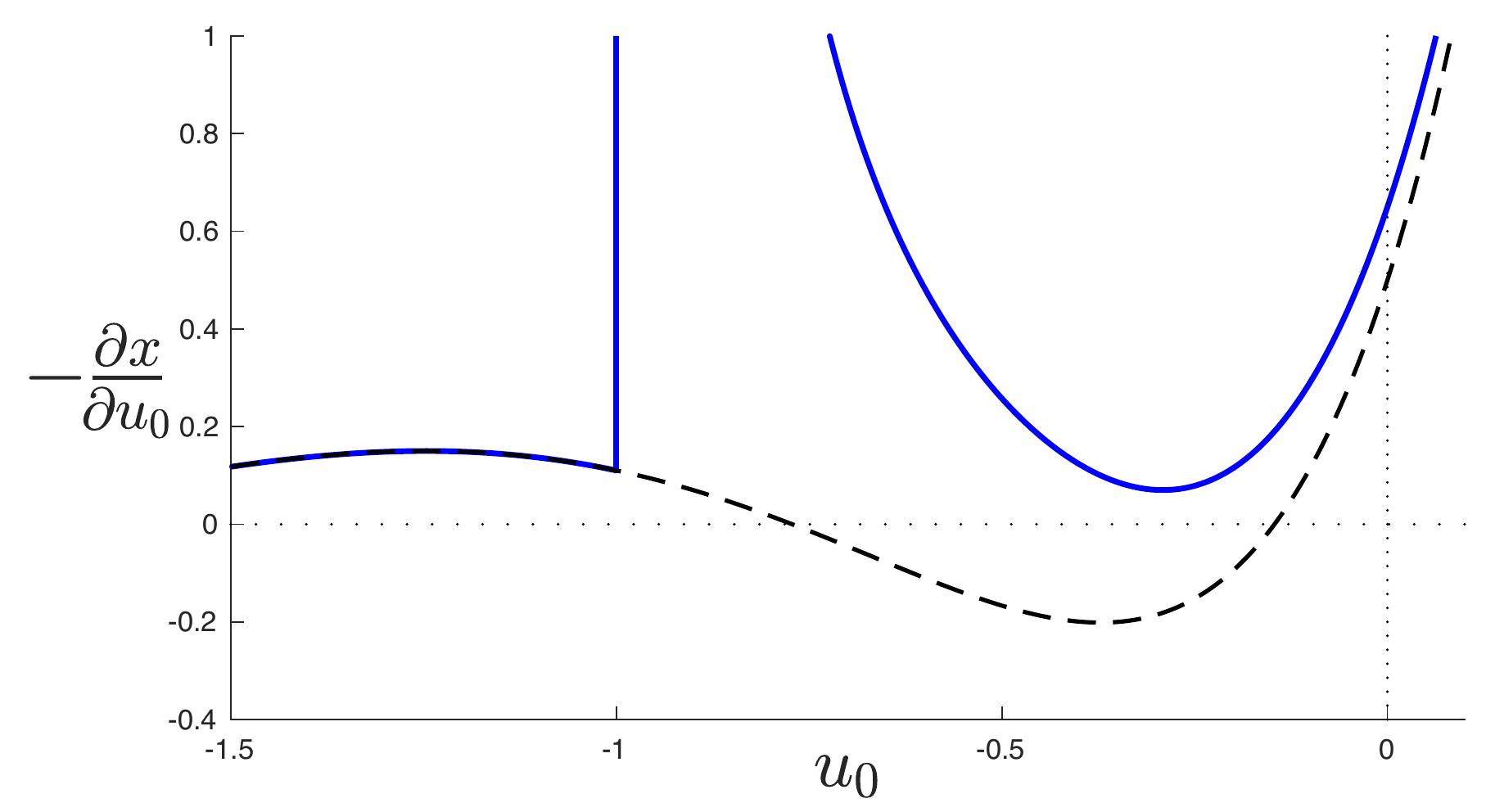}
\caption{\label{fig:JT-zeros2} Same subjects as figure~\ref{fig:JT-zeros} but for ${\tilde \Gamma}{=}0.1$.  In this case, two zeros disappear and so the $u_0(x)$ curve has no folds before it asymptotes to $u_0=\sigma$ as $x{\to}{+}\infty.$}
\end{figure}
\begin{figure}[b]
\centering
\includegraphics[width=0.45\textwidth]{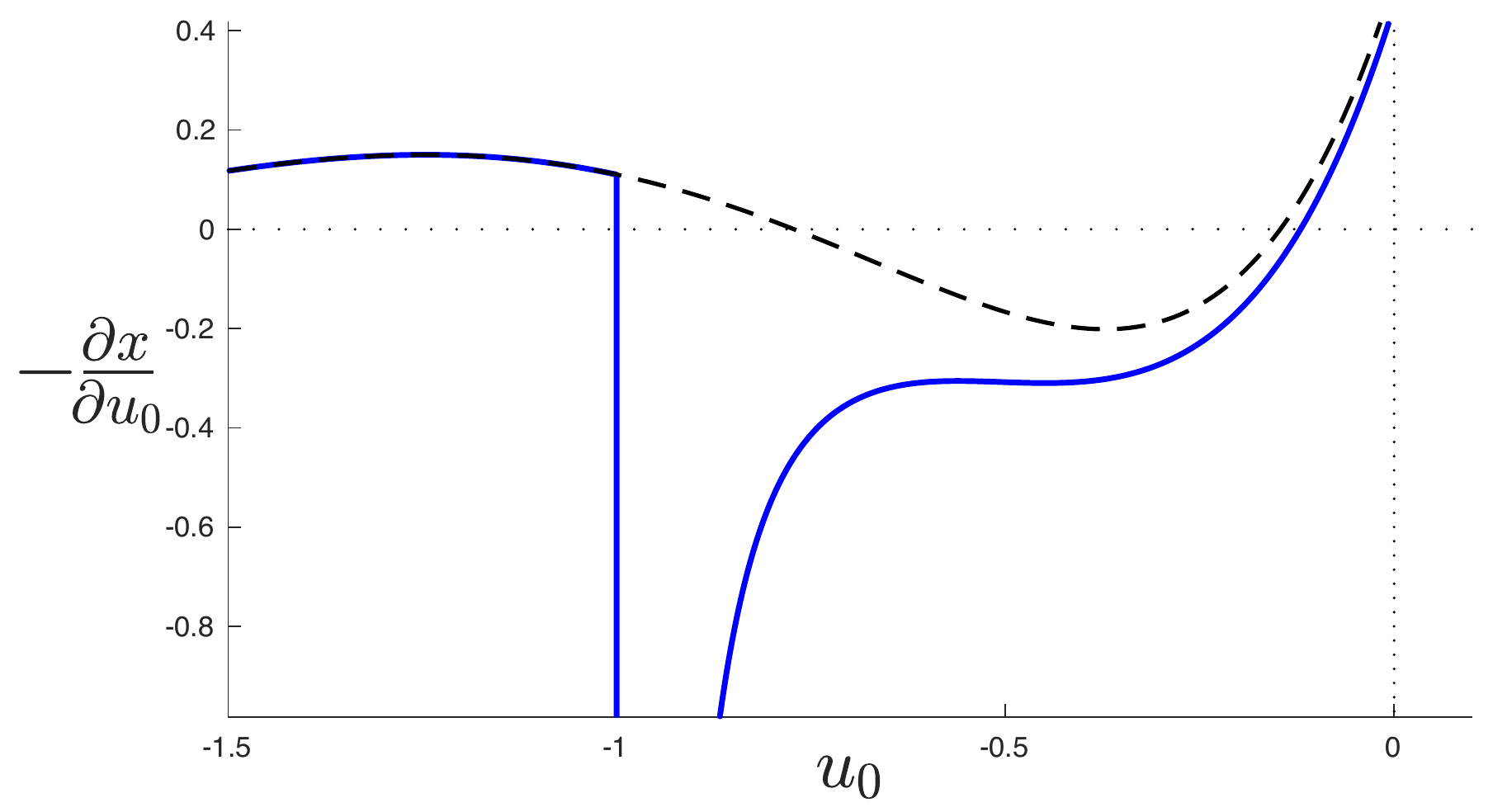}
\caption{\label{fig:JT-zeros3} Same subjects as figure~\ref{fig:JT-zeros} but for ${\tilde \Gamma}{=}{-}0.1$.  In this case, one zero disappears, and so the $u_0(x)$ curve  folds once and then asymptotes to $u_0=\sigma$ as $x{\to}{-}\infty.$}
\end{figure}
It is interesting to look briefly at negative $\tilde\Gamma$ too. Generically it sends the $u_0(x)$ curve out to $x=-\infty$, and  now the original zeros of $u_0(x)$ are deformed in such a way that one happens further to the right. Perturbatively this fits with the idea that the brane is now attractive, and pulls the energies to the left in the spectrum, enhancing the potential of instabilities in the $E<0$ regime since there will now be (perturbatively) a new turning point (and instanton) to the right of  the basic $E=-\frac14$ point. Of course, the whole curve for this case is, according to the consistency conditions described so far, not a good starting point for a non-perturbative completion since it is always folded. The case of ${\tilde\Gamma}{=}-0.1$ and $\sigma{=}-1$ is shown in figure~\ref{fig:JT-zeros3}.  As a final remark for negative $\tilde\Gamma$, note that the deformation can be so strong that $u_0(x)$ drops below zero inside the Fermi sea region of $x<0$. This will result in a solution that is also perturbatively problematic.

Figure~\ref{fig:u-vs-x-at-gamma} shows the example of $u_0(x)$ for the above choices of parameters: ${\tilde \Gamma}{=}{\pm}0.1$, $\sigma{=}{-}1$. The black curve (right, see inset) is for positive~$\tilde\Gamma$ and the red (left) is for negative~$\tilde\Gamma$.

\begin{figure}[t]
\centering
\includegraphics[width=0.48\textwidth]{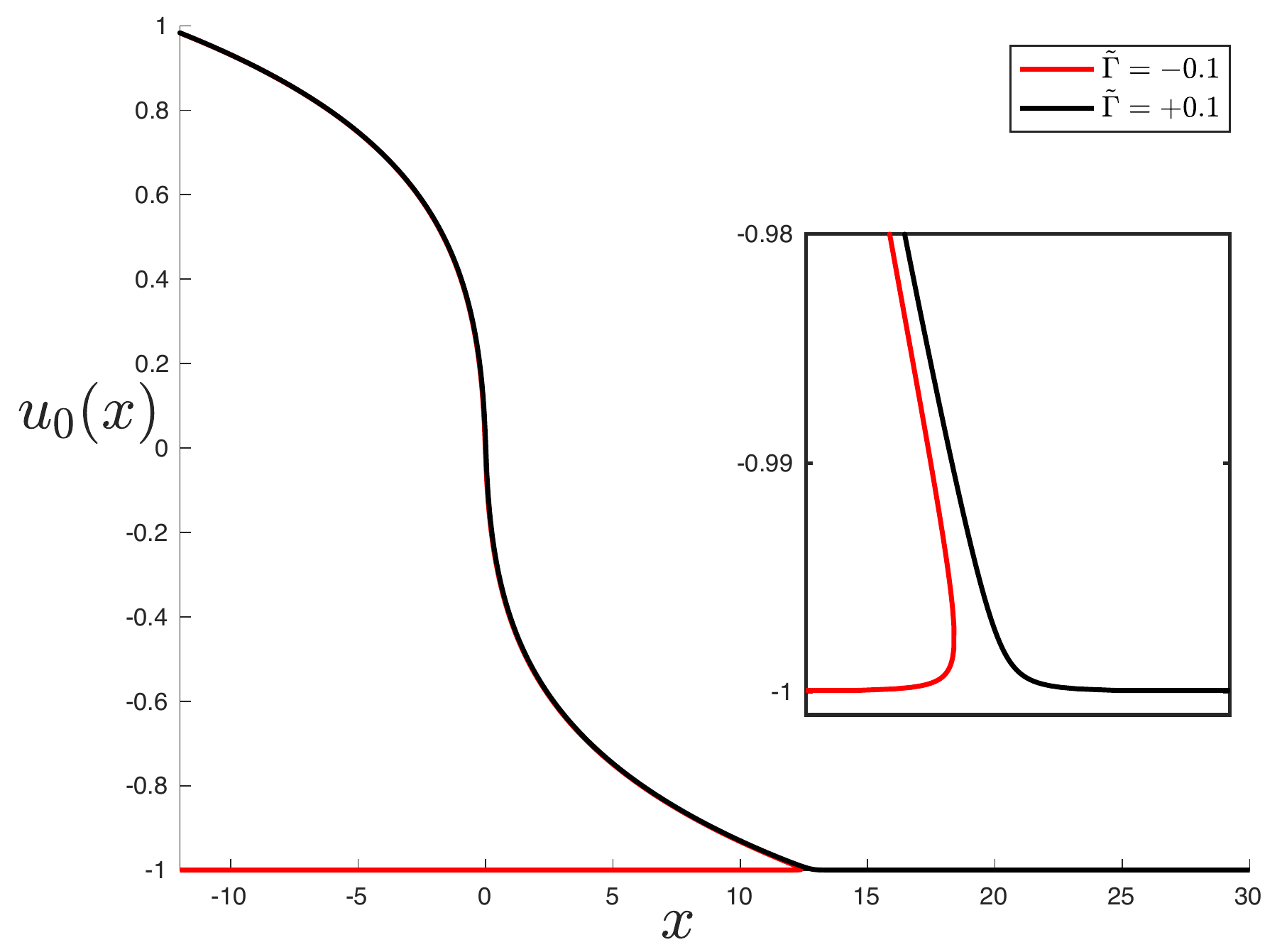}
\caption{\label{fig:u-vs-x-at-gamma} An example of a $u_0(x)$ curve that yields a consistent starting point for a non-perturbative completion, as a result of deformation by a large number of background D-branes. A lowest energy of $\sigma{=}{-1}$ has become accessible. }
\end{figure}

In summary, the addition of background D-branes fits naturally into the consistency conditions. Having a large number of them allows for new non-perturbative completions where the spectrum can go significantly deeper into the negative $E$ regime than is possible without them.\footnote{It should be noted here that a choice was made to place the D-branes at $\sigma$. Any number of them can be introduced, associated with arbitrary ``positions''   along the spectrum.  The  interest here was the extreme end, $\sigma$, of the spectrum, and also the effects of having them coincide, and this is captured succinctly in the string equation~(\ref{eq:big-string-equation}). It would be interesting to see if variants of the string equation can be derived that describe both an endpoint spectrum $\sigma$ {\it and} large collections of branes at other values of $E$.}

\section{Dynamics and Symmetry}
\label{sec:dynamical-boundaries}
The last two sections uncovered a {\it family}  of non-perturbative completions of JT gravity, parameterized by $\sigma\in(u_{\rm t}, 0)$, where $u_{\rm t}{=}{-}(j_{01}/2\pi)^2{\simeq} -0.1464898$.\footnote{As mentioned at the end of Subsection~\ref{sec:consistency-conditions-sub}, this is the complete range of solutions available if the solutions of the string equation have a continuous $\hbar{\to}0$ limit, but it seems that the range can be widened by including solutions that do not have this property.} If $\Gamma$ background D-branes are turned on, they can extend this range of available $\sigma$ even more. That it is a continuous family of completions suggests that they are all connected in some deeper way, and the ability to naturally include D-branes, for which $\sigma$ is their boundary cosmological constant, supports that suggestion. Many aspects of this has already been perturbatively understood in string theory terms in older literature~\cite{Dalley:1992br,Johnson:1992wr,Johnson:1994vk}, although the present context inserts a new, richer flavour into the story. It is sensible to switch to the string theory language for now, since it is helpful. 

The parameter $\sigma$ is a non-perturbative piece of the coefficient of a boundary operator 
that measures loop length. 
Such an operator can always be turned on in the theory, whether
there be background D-branes or not. But when they are
turned on in the manner described in the previous section, then
$\sigma$ is the piece of the boundary cosmological constant associated to
their ``world volume".
Perturbatively  a boundary cosmological constant has the special feature that it is
``redundant", in that it can be accounted for in terms of a mixture
of closed string operators~\cite{Martinec:1991ht}. Such a combination is present here, and it will have an additional mixture with  $\sigma$, as will be precisely explained shortly after some groundwork is laid.

The closed string operators, denoted ${\cal O}_k$, have coefficients denoted $t_k$, but in this section a shift of normalization will be adopted in order to fit with a convention better adapted  to the discussion to come.  The  relation to the $t_k$ of earlier sections  is best stated by redefining the object ${\mathcal R}$ as follows:
\be
\label{eq:new-arkay}
 {\mathcal R} \equiv\sum_{k=1}^\infty\left(k+\frac12\right)t_kR_k+x\ ,
\ee
where $R_k$ is a Gel'fand-Dikii polynomial, but in a
different normalization than used in Section~\ref{sec:string-equations}.
Here, they are given by setting $R_0=2$, and the others are obtained using
the recursion relation 
\be
\label{eq:kdv-recursion-operator}
\hbar R^\prime_{k+1} =\left[ \frac{\hbar^3}{4}\frac{\partial^3}{\partial x^3}-\frac{\hbar}{2}u^\prime-\hbar u\frac{\partial}{\partial x}\right] R_k\ ,
\ee
with the condition that they vanish at $u=0$ for $k>0$. This new normalization of the $t_k$ means a different expression for the JT gravity point than given in equation~(\ref{eq:teekay}), that is closer (up to factors of two conventions) to the expression derived in the  topological gravity approach of ref.~\cite{Dijkgraaf:2018vnm}.

The insertions of ${\cal O}_k$ are well known~\cite{Gross:1990aw} to be equivalent to
changing $u(x)$ according to the KdV flows:
\be
\label{eq:KdV}
\frac{\partial u}{\partial t_k} = -\frac{\partial}{\partial x} R_{k+1}[u]\ ,
\ee

It is useful to observe from the KdV equation the following
relation between the scaling dimensions of $u$ and the
variables $x$ and $\{t_k \}$: $ k [u] = [x]-[t_k]$. In
particular, since the zeroth flow implies $x =-t_0$, they
have the same scaling. Alternatively, using the recursion
relation to rewrite the right hand side implies
$(k-1) [u] =3 [x] -[t_k]$. Hence, $[x] =-\frac12[u]$,
and $[t_k] =-(k+\frac12)[u]$.
This means that if $u$ transforms nicely under some scaling by $\Delta$:
\be
\label{eq:scalings}
\Delta u(x;t_k) = u(\Delta^{-\frac12}x,\Delta^{-(k+\frac12)}t_k)\ ,
\ee
which implies:
\be
u+\frac12x\frac{\partial u}{\partial x}+\sum_{k=1}^\infty\left(k+\frac12\right)t_k\frac{\partial u}{\partial t_k}=0\ .
\ee
Using KdV and the recursion relation gives
\be
u+\frac12 xu^\prime-\sum_{k=1}^\infty\left(k+\frac12\right)t_k\left\{\frac{\hbar^2}{4}R_k^{\prime\prime\prime}-\frac12 u^\prime R_k-uR_k^\prime\right\}\! = 0\ ,
\ee
{\it i.e.,}
\be
\label{eq:scaling-2}
 u{\mathcal R}^{\prime} +\frac12 u' {\mathcal R} - \frac{\hbar^2}{4} {\mathcal R}^{'''}=0\ ,
\ee
recalling that, here, ${\mathcal R}$ is given in equation~(\ref{eq:new-arkay}).
Multiplying equation~(\ref{eq:scaling-2}) by ${\mathcal R}$ and integrating once with respect to $x$ gives the string equation~(\ref{eq:big-string-equation}) with $\sigma{=}0$ and
the integration
constant is $\hbar^2\Gamma^2$.
In other words, the string equation used in previous sections to discover non-perturbative completions of JT follows from
scale invariance and assuming KdV. 

It is straightforward to incorporate $\sigma$ here.  Its presence  in the string equation promotes (\ref{eq:scaling-2}) to:
\be
 (u-\sigma){\mathcal R}^{\prime} +\frac12 u' {\mathcal R} - \frac{\hbar^2}{4}{\mathcal R}^{'''}=0\ .
\ee
If read as a scaling equation again, it
says $\sigma {\mathcal R}^\prime=-\sigma\frac{\partial u}{\partial\sigma}$, since $u$ and $\sigma$
have the same dimension.
Hence both 
equations read:
\bea
\label{eq:thing1-and-thing2}
 -\frac{\partial u}{\partial\sigma}&=& {\mathcal R}^\prime \equiv {\cal D}{\mathcal R} \ , \nonumber\\
-\sigma \frac{\partial u}{\partial\sigma}  &=& u{\mathcal R}^{\prime} +\frac12 u' {\mathcal R} - \frac14{\mathcal R}^{'''} \equiv ({\cal L}{\cal D}^{-1}){\cal D} {\mathcal R}\ ,
\eea
where $D{\equiv}\hbar\partial_x$ and ${\cal L}$ is the recursion operator~(\ref{eq:kdv-recursion-operator}). In fact, a semi-infinite tower of equations can be generated~\cite{La:1990sn} by  acting with $({\cal L}{\cal D}^{-1})^{n+1}$, for higher integer~$n$. More will be said about this  below.

The key observation to make next is that the first relation related derivatives  with respect to $\sigma$ to derivatives with respect to the various $t_k$. Acting on loop operators to see precisely how they behave when thus differentiated results in:
\bea
&&\left(\frac{\partial}{\partial\sigma}+\sum_{k=1}^\infty \left(k+\frac12\right) t_k\frac{\partial}{\partial t_{k-1}}\right)\langle w(\ell)\rangle = \ell\langle w(\ell)\rangle\nonumber \\&&\hskip 5cm \equiv \langle {\cal O}_B w(\ell)\rangle\ ,
\eea
showing the particular combination of $\sigma$ and the closed string operator coefficients $t_k$ that combine to act as a boundary cosmological constant, coupling to the length operator ${\cal O}_B$.


The parameters $t_k$ that bring in the various minimal model components had their values  set according to equation~(\ref{eq:teekay}) in order to yield the leading Schwarzian JT gravity result. So for those fixed $t_k$, different $\sigma$ are indeed inequivalent non-perturbative completions. However,  deformations of the gravity background can be interpreted~\cite{Johnson:2020lns,Rosso:2021orf} as changing the $t_k$. The results just derived connecting changes in $u(x)$ due to the $t_k$ to changes due to $\sigma$ shows that $\sigma$ can be generally thought of as a dynamical coupling too. This makes sense since if a given set, $\{ t_k\}$ were to change, the value of $\sigma$ used in defining the non-perturbative completion should be expected to change in response.\footnote{A precursor of this was seen in this context in ref.~\cite{Johnson:2020lns}.}  The language to use for all these potential changes to $\{t_k;\sigma\}$ is simple to state, since it is all incorporated into how the  function $u(x)$ adjusts itself: Changes of $u(x)$ under $t_k$ are given in terms of the KdV flows, and there is an additional flow for $\sigma$ as given in the first relation in~(\ref{eq:thing1-and-thing2}).  Writing $u{=}-\hbar^2\partial^2_x\log \tau $, that expression can be written as (after two $x$ integrations):
\be
\left(L_{-1}-\frac{\partial}{\partial\sigma}\right)\cdot\tau=0\ ,
\ee
where the operator
\be
L_{-1}\equiv\sum_{k=1}^\infty \left(k+\frac12\right)t_k\frac{\partial}{\partial t_{k-1}}+\frac{x^2}{2\hbar^2}\ .
\ee
Similarly, the second relation in~(\ref{eq:thing1-and-thing2}), which is  the (once differentiated) string equation~(\ref{eq:big-string-equation})) can be massaged into the following form:
\be
\left(L_0-\sigma\frac{\partial}{\partial\sigma}\right)\cdot\tau = 0\ ,
\ee
where
\be
L_{0}\equiv\sum_{k=0}^\infty \left(k+\frac12\right)t_k\frac{\partial}{\partial t_{k}}+\frac{1}{16}\ .
\ee
Almost all of the constants of integration must vanish for closure of the algebra discussed below, while the $L_0$ eigenvalue is set by  the underlying conformal field theory structure (see below).  As observed under equation~(\ref{eq:thing1-and-thing2}), further action with higher powers of the recursion operator ${\cal L}{\cal D}^{-1}$ gives more relations, and after $x$-integrating twice and fixing some constants as before, they can be written as 
\be
\left(L_n -\sigma^{n+1}\frac{\partial}{\partial\sigma}\right)\cdot\tau = 0\ ,
\ee
where, for $ n\geq 1$:
\be
\label{eq:twisted-boson}
L_n\equiv \sum_{k=0}^\infty \left(k+\frac12\right)t_k\frac{\partial}{\partial t_{k+n}}+\frac{\hbar^2}{8}\sum_{k=1}^n\frac{\partial^2}{\partial t_{k-1}\partial t_{n-k}}\ .
\ee
These are  Virasoro constraints~\cite{Dijkgraaf:1991rs,Fukuma:1991jw}, but of a modified form first explored in refs.~\cite{Dalley:1991vr,Dalley:1991xx,Johnson:1992wr,Johnson:1994vk}, that act on  the  $\tau$-function~\cite{Date:1981qy} that $u$ defines, forming the algebra $[L_n,L_m]=(n{-}m)L_{m+n}$. (A modification when $\Gamma$ background D-branes is turned on will be discussed shortly.) The operators~(\ref{eq:twisted-boson}) have an interpretation as the modes of the stress tensor of a $\mathbb{Z}_2$--twisted boson $\phi(z)$ on (yet another) auxiliary space, essentially an extension of $E$ to the complex $z$ plane (the natural home of the matrix model's ``spectral curve''). The $1/16$ in the $L_0$ constraint is consistent with the twist.

 The first two constraints express invariance under certain symmetries: translations and Galilean transformations for $L_{-1}$, incorporating a shift of  $\sigma$ and the mixing together of the $t_k$ (seen in the boundary operator), and scalings for $L_0$ of the form  seen in equation~(\ref{eq:scalings}). The whole family expresses the  diffeomorphisms of a line, now with a movable boundary at $\sigma$. That line is the scaled Dyson gas itself, {\it i.e.} the spectrum on $[\sigma,+\infty)$.

Now it is clear from yet another perspective why the string equation~(\ref{eq:big-string-equation}) is the one that fully defines the model by giving a sensible $u(x)$ non-perturbatively. For a given consistent one-cut solution needed for JT gravity, the spectrum lives on the half-line, with some fixed  lower bound $\sigma$. This means translation invariance is broken and therefore the $L_{-1}$ string equation~(\ref{eq:simple-string-equation}) is not available. Scale invariance is preserved, however, and so the~$L_0$ equation applies.

Finally, this algebraic structure all readily extends~\cite{Dalley:1992br} to the case of having added $\Gamma$ background D-branes in the manner done in Section~\ref{sec:D-branes}. In fact, it was shown in ref.~\cite{Johnson:1994vk} that there is a very simple way to construct it. Starting with the purely closed string system with just the $t_k$, the following shift of the couplings adds $\Gamma$ D-branes with boundary cosmological constant $\sigma$:
\be
\label{eq:coupling-shift}
t_k\to t_k+2\hbar\Gamma \frac{\sigma^{-\left( k+\frac12\right)}}{\left(k+\frac12\right)}\ ,
\ee
resulting in the following Virasoro constraint operators that act on a new $\tau$-function representing the  string theory with open string sectors (D-branes):
\be
\left(L_n - (n+1)\frac{\Gamma^2}{4}\sigma^n-\sigma^{n+1}\frac{\partial}{\partial\sigma}\right)\cdot\tau = 0\ ,
\ee
Following the algebra through carefully~\cite{Johnson:1994vk} shows that the new system can be interpreted as adding (in the twisted boson language),  at position $z{=}\sigma$,   a vertex operator $V(\sigma){=}:\e^{-\frac{\Gamma}{\sqrt{2}}\phi(\sigma)}:$ that has weight $\Gamma^2/4$, accounting for the middle term in the new Virasoro operators.\footnote{It was noted in ref.~\cite{Johnson:2004ut} that the cases of $\Gamma{=}\pm\frac12$ are interesting because the $\frac{1}{16}$ of $L_0$ gets cancelled. Since in the current context those are special instances of supersymmetric JT gravity models (but with $\sigma{=}0$, $\mu{=}1$, and a different recipe for the $t_k$~\cite{Johnson:2020heh,Stanford:2019vob}), it would be interesting to explore this further.} While the discussion in that work considered only perturbation theory, it is clear that the same structure persists here beyond perturbation theory, and the choice has been made to place the branes at~$\sigma$ which is a non-perturbative parameter here, as has been discussed.

Before ending this section, it should be noted that there is an additional  criterion that  distinguishes different non-perturbative completions. As noted above, the  framework  is organized by the KdV flows, which are evident perturbatively in the basic Hermitian matrix model. 
Moreover,  it is possible to derive the central string equation~(\ref{eq:big-string-equation}) as a consequence of just the KdV flows~(\ref{eq:KdV}) combined with the scale invariance of $u(x,t_k;\sigma)$ (and hence of the Dyson gas) described in equation~(\ref{eq:scalings}). So if the full $u(x)$  implied by any non-perturbative completion {\it does not} satisfy the string equation, then it non-perturbatively violates either the KdV flows or scaling.\footnote{This re-animates an old idea of refs.~\cite{Dalley:1991qg,Dalley:1991vr} that was used quantitatively in ref.~\cite{Ambjorn:1991km}.} 
%
%
%
%
\section{Closing Remarks}
\label{sec:discussion}

There is a tacit assumption throughout this paper that it is not merely a  coincidence that a double scaled Hermitian matrix model captures JT gravity perturbatively. Given that, a search for a non-perturbatively stable completion of the physics should stray as little as possible from this perturbative setting, doing the minimal relaxation of the assumptions made. 

The  results of this paper arose from following the physics in this spirit. Seeking a perturbative Hermitian matrix model description of JT gravity to leading order  leads  to the requirement of a single cut configuration filling the region $E\,{\geq}\,0$. However, beyond perturbation theory this configuration is not a good solution of the unconstrained Hermitian matrix model. This is signaled by an instability   to developing other cuts,  seen semi-classically at some  low enough $E<0$. This suggests a solution to the problem where  the  lowest energy, $\sigma$, of the spectrum, occurs away from the where the instability sets in. The original non-perturbative proposal of ref.~\cite{Johnson:2019eik} was such a solution, with $\sigma{=}0$. This paper shows that  $\sigma$ can also take other values, and is a natural   non-perturbative parameter of the theory.  Put differently, the statement is:

\begin{itemize}[label={\raisebox{2pt}{\tiny\textbullet}},leftmargin=*,itemsep=-0.5ex]  
\item{\it  JT gravity is fully defined as an ensemble of random Hermitian matrices restricted to lowest eigenvalue $\sigma$.}  
\end{itemize}

\noindent A natural concern is that $\sigma$ seems arbitrary, but it is not. An analysis shows that it is  tightly constrained  by simple consistency conditions to have a narrow range of values given in equation~(\ref{eq:broader-defintion}) (Section~\ref{sec:D-branes} showed that the range can be enlarged or reduced with the addition of background D-branes). Crucially, the Hermitian matrix model supplies a string equation~(\ref{eq:big-string-equation}) that contains  the needed stable non-perturbative descriptions, but only for that special range, while at the same time reproducing perturbation theory.  Moreover $\sigma$, and the fact that it is continuous, was seen to be a natural component of the string theory language that organizes the entire family of models (which can include background D-branes) of which this system is part.

As an aside, a key point  is that the natural setting here is entirely within Hermitian matrix models. However, it {\it is} possible to  interpret the more general string equation~(\ref{eq:big-string-equation}) as a deformation of a complex matrix model ensemble, and therefore the individual minimal model components as deformed Type~0A minimal models. They are combined in a manner~(\ref{eq:teekay}) that is very different from the JT supergravity points described~\cite{Johnson:2020heh} using complex matrices however. Nevertheless some local features can be made to emerge by experimentation, as done by tuning $\mu$ from zero to larger positive values in ref.~\cite{Johnson:2019eik}, bringing out the Bessel behaviour characteristic of SJT physics. (See  ref.~\cite{Rosso:2021orf} for a phase transition where $\mu$ {\it necessarily} changes sign as part of a family of SJT deformations.)  Overall it is worth remarking that there could be something to be learned by thinking further along these lines, where JT gravity becomes a special point in a space of theories that also includes the JT supergravity theories.\footnote{CVJ thanks Edward Witten for a conversation on this point.}

Going back to the main discussion, a major benefit of using this non-perturbative framework is the fact that is allows for explicit computation of physical quantities, and, crucially, seamlessly connects to the perturbative description, which is straightforwardly translated into the perturbative language used in Saad, Shenker and Stanford~\cite{Saad:2019lba}, connecting to  perturbative gravitational computations. Hence, another goal of this paper was to carefully lay out, for future use,  how the framework pieces together the various classical, perturbative, semi-classical, and fully non-perturbative regimes.  In this regard, it was emphasized that the many-body fermion description of the matrix model organizes the framework beautifully, with the Fermi sea region ($x{\leq}\mu$) being all that is needed for perturbation theory, and  the ``trans-Fermi'' region $x{>}\mu$ being  crucial for the non-perturbative story. 

Imagine that there exists some other (not necessarily Hermitian matrix model) approach to the perturbative description of JT gravity. 
While it might not be straightforward to do, it is implicit that {\it any}  such description can be cast into the language used here,  amounting to a perturbative description of the function $u(x){=}\sum_{g=0}^\infty u_g(x)\hbar^{2g}$ for $x$ in the Fermi sea regime $x\leq\mu$. All the $u_g(x)$ can be uniquely described by the Hermitian matrix model (see~(\ref{eq:leading-u}) for $u_0(x)$ and then expand~(\ref{eq:simple-string-equation}) in $\hbar/x$ for the others), but it is possible to imagine that this alternative approach  computes them in a very different manner.  

A key suggestion of this paper is the next logical step: Any non-perturbative completion of that alternative approach amounts to an extension of $u(x)$ into the trans-Fermi regime $x{>}\mu$. Different completions imply different behaviour for $u(x)$ in the trans-Fermi regime, and so can be quantitatively compared to each other, as was done  for the examples explored here. It would be interesting to see if the non-perturbative suggestions of Gao, Jafferis, and Kolchmeyer~\cite{Gao:2021uro} (following ref.~\cite{Saad:2019lba}) and possibly other proposed completions to come, can be shown to imply a $u(x)$ for $x{>}\mu$ whose features can then be compared to the definitions used here.  Even some partial results for the features of such a $u(x)$ would be interesting to explore, since they should yield a quantifiable imprint on the physical quantities.

It is widely expected  that 
JT gravity presumably embeds  into a more complete theory. Perhaps it arises as part of a D-brane configuration in some higher dimensional gravity setting that builds a charged or rotating black hole solution. From the results seen here, evidently there is a family of possible embeddings, parameterized by a natural stringy parameter $\sigma$ and the work here has supplied constraints on what values of $\sigma$ are possible. An intriguing prediction is that within this more complete setting, if~$\sigma$ evolves to move outside the stable range established by the methods of this paper,  there will be a phase transition in the full theory. The multi-cut phase may well be part of its description.

 \begin{acknowledgments}
CVJ  thanks    Felipe Rosso  for many questions and helpful remarks, and  members of the research groups at Princeton University and the Institute for Advanced Study for comments and questions, especially Robbert Dijkgraaf, Juan Maldacena, Joaquin Turiaci, Herman Verlinde, and Edward Witten. CVJ also thanks  the  US Department of Energy for support under grant  \protect{DE-SC} 0011687, and  Amelia for her support and patience.    
\end{acknowledgments}

\bibliographystyle{apsrev4-1}
\bibliography{Fredholm_super_JT_gravity1,Fredholm_super_JT_gravity2}

\end{document}